# Flow coupling between active and passive fluids across water–oil interfaces

Yen-Chen Chen[1], Brock Jolicoeur[2], Chih-Che Chueh[3], and Kun-Ta Wu[1,2,4,*]

[1]Department of Mechanical Engineering, Worcester Polytechnic Institute, Worcester, Massachusetts 01609, USA
[2]Department of Physics, Worcester Polytechnic Institute, Worcester, Massachusetts 01609, USA
[3]Department of Aeronautics and Astronautics, National Cheng Kung University, Tainan 701, Taiwan
[4]The Martin Fisher School of Physics, Brandeis University, Waltham, Massachusetts 02454, USA
[*]Corresponding: kwu@wpi.edu

**Abstract**
Active fluid droplets surrounded by oil can spontaneously develop circulatory flows. However, the dynamics of the surrounding oil and their influence on the active fluid remain poorly understood. To investigate interactions between the active fluid and the passive oil across their interface, kinesin-driven microtubule-based active fluid droplets were immersed in oil and compressed into a cylinder-like shape. The droplet geometry supported intradroplet circulatory flows, but the circulation was suppressed when the thickness of the oil layer surrounding the droplet decreased. Experiments with tracers and network structure analyses and continuum models based on the dynamics of self-elongating rods demonstrated that the flow transition resulted from flow coupling across the interface between active fluid and oil, with a millimeter–scale coupling length. In addition, two novel millifluidic devices were developed that could trigger and suppress intradroplet circulatory flows in real time: one by changing the thickness of the surrounding oil layer and the other by locally deforming the droplet. This work highlights the role of interfacial dynamics in the active fluid droplet system and shows that circulatory flows within droplets can be affected by millimeter–scale flow coupling across the interface between the active fluid and the oil.

**Introduction**
Active fluids flow without external energy input owing to force generated by active entities that consume local fuel to generate kinetic energy.[1-10] Active fluids can self-organize into circulatory flows that are sensitive to confinement shape.[11-25] However, little is known about the role of boundary conditions in the self-organization of confined active fluids, especially fluid boundaries such as water–oil interfaces. Fluid boundaries are known to induce the coupling of the fluid dynamics on both sides of the boundary owing to hydrodynamic coupling.[26,27] For example, when active fluid is confined in a droplet immersed in liquid crystal, the liquid crystal develops oscillating rings surrounding the droplet,[28-31] which indicates that the active fluid alters the passive fluid configuration on the opposite side of the boundary. However, the principles underlying such coupling and the associated fluid mechanics have not yet been elucidated. Understanding such coupling dynamics is essential to unravel the dynamic role of fluid boundaries at interfaces between passive and active fluids and their impact on active fluid flows. Here, we investigated the hydrodynamic coupling between active and passive fluids in a water-in-oil active fluid droplet system with an approach that combines experiments and modeling. In our experiments, we confined an active fluid in a water-in-oil droplet that was compressed into a cylinder-like shape[32] and characterized the flow coupling between active fluid and oil (passive fluid) near water–oil interface. We focused on how active fluids drive the oil through interface and on how the oil configuration can, in return, influence the self-organization of active fluid. We also determine the characteristic length scale of this active–passive fluid coupling. To gain deeper insight into our experimental results, we developed a continuum complex fluid simulation based on established active fluid models[26] and explored methods of directing active fluid flows with novel millifluidic devices that can manipulate interfacial dynamics and droplet shapes in real time.

## Results and Discussion

**Kinesin-driven, microtubule-based active fluid.** To investigate coupling between active and passive fluids across a fluid boundary, we selected microtubule–kinesin complex active fluid because it is tunable and reproducible and has established characterizations and models to serve as references.[33-37] Microtubule–kinesin active fluid is driven by extensile microtubule bundles that are assembled by depletion and extended by kinesin motor clusters (Fig. 1a).[38] The bundles repeatedly extend, brake, and anneal, resulting in an active gel network whose structure is constantly rearranged.[32] The rearranging active gel stirs the surrounding liquid (96% water), which causes flows. In this paper, we study the fluid dynamics of this active fluid when confined in a fluid cylindrical boundary.

**Confining active fluid in an immobilized active droplet.** To create a fluid cylindrical boundary, we confined the active fluid in water-in-oil droplets compressed by plates at the top and bottom. The combination of compression and interfacial tension deformed the droplets into cylinder-like shapes.[39] However, the compressed droplets were motile,[26,32,40] so we curved the top plate into a dome-like shape to fix the droplet at the dome center and thus allow the long-term observation of intradroplet flows (Fig. 1b and Supplementary Discussion S1). The flows within the compressed droplets self-organized into droplet-wise circulatory flows (Supplementary Video S1)[13] whose circulation depended on the thickness of the oil layer surrounding the droplet (Fig. 1c). The intradroplet circulatory flows were more pronounced with a thicker layer of oil but suppressed with a thinner layer of oil (Supplementary Video S2).

To quantify the coherence of the circulatory flows, we added fluorescent tracer particles to the active fluid. We defined the circulation order parameter (COP) as the average fraction of the azimuthal component in the flow velocity: $\text{COP} \equiv \langle v_{i,\theta}/|\boldsymbol{v}_i|\rangle_i$, where $v_{i,\theta}$ is the azimuthal component of velocity $\boldsymbol{v}_i$ of the $i^{\text{th}}$ tracer, and $\langle\ \rangle_i$ indicates averaging of the tracer particles.[11] A COP of 1 indicates perfect counterclockwise circulation, a COP of −1 indicates perfect clockwise circulation, and a COP of 0 indicates chaotic flow. The COP analyses revealed that, in droplets with a thick oil layer, circulatory flow developed over the first hour and remained at a steady state for several hours. By contrast, in droplets with a thin oil layer the COP fluctuated (Fig. 1d). These results demonstrate that self-organization of active fluid confined in fluid boundaries is sensitive to parameters outside the boundaries.

**Intradroplet circulatory flows depended on the geometries of droplets and oil.** To investigate how the intradroplet flows were influenced by the shapes of the droplet and oil, we first immersed compressed active droplets with the same height ($h = 1$ mm) but different radii ($r = 0.5$–5 mm) in oil baths ($R = 5$ mm) and measured the COP (Fig. 2a). Our measurements revealed that circulatory flow developed when the droplet radius was smaller than a critical threshold ($r_c \approx 2.4$ mm); enlarging the droplet above this limit suppressed the circulatory flow (Fig. 2a). Moreover, we found that this critical radius was reduced ($r_c \approx 1.4$ mm) when the oil bath was smaller ($R = 3.5$ mm; Fig. 2b). These results highlight the role of oil configuration in the formation of intradroplet circulatory flows.

**Circulatory flows within droplets depended on the oil layer thickness.** To characterize the influence of the oil configuration on intradroplet flows, we fixed the droplet shape ($r \approx 2.4$ mm, $h = 2$ mm), varied the size of the oil bath ($R$) or the thickness of the oil layer surrounding the droplet ($\Delta \approx 0$–11 mm) and measured the COP as a function of oil layer thickness (Fig. 3a). Our measurements revealed that the intradroplet flows were sensitive to the oil configuration when the oil layer was thinner than a critical thickness ($\Delta_c \approx 4$ mm); oil layers thicker than this limit did not influence intradroplet flows. We hypothesized that the critical thickness of the oil layer originates from active fluid–oil coupling near the interface, which should not be sensitive to droplet shape.[40] To test this hypothesis, we repeated the measurements with droplet shapes with different $r$ and $h$ values (Figs. 3b–d). The results showed that across our explored droplet geometries, the critical thickness was similar. When the oil layer was thinner than this thickness, the COP varied, whereas when the oil layer was thicker than the critical thickness, the circulatory flows remained intact. The existence of this universal critical thickness implied that the intradroplet active fluid was coupled to the oil within the critical thickness from the interface, which is consistent with our hypothesis.

To investigate how this coupling affected the net flow rates of intradroplet circulation, we analyzed the flow profiles of averaged azimuthal velocities (Figs. 3e–h) and found that the intradroplet flows that had a higher COP flowed more coherently in the azimuthal direction so had a higher net azimuthal flow rate. Conversely, the flows that had a lower COP flowed more chaotically and so had a lower net flow rate. A slower flow rate does not indicate a slower flow speed; rather, it indicates that the flows are more chaotic and thus the positive (counterclockwise) and negative (clockwise) azimuthal flows cancel each other out. The flow speed of intradroplet active fluid was independent of oil layer thickness. Our analysis revealed that thickening the oil layer irregularly supported, suppressed, or even ceased the net azimuthal flows, which agreed with our COP data (Figs. 3a–d). This irregularity revealed a nonlinear influence of the active fluid–oil coupling on circulatory flows. Despite the complex nature of the active fluid–oil coupling, our data demonstrated that the oil configuration could direct the self-organization of intradroplet active fluid through the water–oil interface.

**The oil layer thickness influenced microtubule network structure in the droplet.** To explore whether the microtubule network structure was affected by the oil configuration, we analyzed confocal microscopy images of the active fluid droplet with the snake algorithm to extract the network structure and reveal the bundle orientational distributions (Figs. 4b–d).[41] Our analyses revealed that when the droplet ($r \approx 2.4$ mm, $h = 1$ mm) developed circulatory flows, the microtubule bundles near the water–oil interface tended to align at an angle ~15° from parallel with the interface (blue solid curve in Fig. 4d). The alignment decreased with increasing distance from the interface (red solid curve in Fig. 4d), and at the droplet center the microtubule bundles were oriented randomly (green solid curve in Fig. 4d). This indicates that circulatory flows within droplets were accompanied by a thin nematic layer of microtubule bundles near the water–oil interface.[11] Furthermore, this formation of a nematic layer was suppressed when the thickness of the oil layer was decreased and the intradroplet circulatory flow was suppressed (dashed curves in Fig. 4d). These results suggest that fluid dynamics in the oil penetrate the water–oil interface and influence the self-arrangement of the microtubule network inside the droplet. This is consistent with our observations that circulatory flows are dependent on the thickness of the oil layer (Fig. 3).

**The oil layer thickness influenced active stress distribution in the droplet.** To gain insight into the impact of oil layer thickness from the perspective of fluid dynamics, we analyzed the time-averaged flow fields, vorticity maps, director fields, nematic order parameter distributions, and active stress distributions in active fluid near the water–oil interface (Fig. 4e–j). We found that, for a droplet immersed in a thicker oil layer ($\Delta = 2.4$ mm), the directors were mostly aligned with the fluid flow and the nematic order parameter decreased with distance from the interface (Fig. 4e&f). This variation in alignment order led to a gradient in active stress near water–oil interface (Fig. 4g). This active stress gradient generated forces that directed the coherent flow (Fig. 4e). Conversely, for the droplet immersed in a thinner oil layer ($\Delta = 1.1$ mm), the directors oriented more chaotically with nearly zero nematic order parameter near the water–oil interface (Fig. 4i). As such, the active stress was uniform (Fig. 4j) and the net flow velocity was nearly zero (Fig. 4h). These results show flow coupling across the interface between active fluid and oil that influences the stress distribution in the active fluid and thus can direct active fluid flows.

**Active fluid in the droplet induced chaotic flows in the oil.** Our data showed that the thickness of oil layer surrounding the droplet influenced the self-organization of the intradroplet active fluid flows, which suggests an interaction between flows in the droplet and flows in the oil. To reveal such an interaction, we monitored the flows in the oil ($\Delta \approx 2.4$ mm) as well as in the droplet ($r \approx 2.4$ mm, $h = 1$ mm). To distinguish the flows in both regimes, we doped the oil with 1-µm tracers and doped the droplet with 3-µm tracers (Fig. 5a) and monitored the tracers for 1 hour. Time-averaged velocity fields and vorticity maps (Fig. 5b) and flow profiles of azimuthal velocities (Fig. 5c) showed no observable net flows in oil in either circulating or noncirculating droplets. However, the absence of net flow does not necessarily imply that the oil is quiescent, as zero net flows can result from chaotic flows whose velocities canceled out over a time average.[11] Flow speed profiles revealed that oil near the interface developed 2- to 4-µm/s chaotic flows that decayed with distance from the interface with a decay length of ~0.5 mm (Fig. 5d). This decay length suggests that the dynamics of the intradroplet active fluid were coupled to the oil near the interface with a millimeter–scale coupling length. This implies that disturbances to the oil (such as stirring) within this coupling range might influence the intradroplet active fluid flows even if the disturbance does not physically contact the droplet. Conversely, a disturbance outside this coupling range might not affect the intradroplet circulatory flows. This suggestion is consistent with our observation that changing the thickness of oil layers within a critical thickness affects the formation of intradroplet circulatory flows and *vice versa* (Fig. 3).

These results characterizing flows in the oil surrounding active fluid droplets are consistent with the model prediction by Young *et al.* that the surrounding oil will remain quiescent when an active droplet develops circulating flows and will be driven to flow when the active fluid flows are noncirculating (such as extensile or quadruple flows).[27] Our data showed that when the droplet was in a circulating state (Fig. 5d, blue curve), flows in the oil only developed near the water–oil interface, but when the droplet was in a chaotic state (red curve) the flows in the oil extended to the container surface ($\rho = 1.1$ mm). However, Young *et al.* predicted that intradroplet circulation would be accompanied by a counter-rotation of active fluid near the water–oil interface, which was observed in bacteria-based active droplets,[14] but our data did not show this counter-rotation (Figs. 3e–h and 5b&c). A possible explanation for this discrepancy is system dimensionality; Young *et al.*'s model and bacteria-based active droplet system were both two dimensional, whereas our active droplet is a three-dimensional system in which confinement by the ceiling and floor might have induced additional friction that inhibited the development of counter-rotation near the interface.

**A continuum simulation qualitatively agreed with experimental outcomes.** Our experimental results show that the coupling of flows within and outside droplets influenced the self-organization of flows within the droplet. To gain deeper insight into this flow coupling, we modeled the active fluid droplet system with an existing active fluid model developed by Gao *et al.* in 2017.[26] We selected Gao *et al.*'s model because, while active fluids have been modeled using swimmer-based simulations[42,43] and continuum equations of mean fields of active particles,[44-50] Gao *et al.*'s framework not only includes multiphase fluids (oil and water) along with associated interfaces, but it also succeeds in describing the self-propelling and self-rotating characteristics of active droplets.[26] Moreover, Gao *et al.*'s model shows that active fluid encapsulated in a water-in-oil droplet can induce flows in the surrounding oil, which was observed in our experiments (Fig. 5).[26] Therefore we adopted Gao *et al.*'s model to test its capability to describe our experimental outcomes.

*Model description.* The model considered two main forces: (1) interfacial tension force from the droplet surface, $\boldsymbol{T} \equiv (\gamma K/\epsilon)\boldsymbol{\nabla}c$, where $\gamma$ is the water–oil interfacial tension; $\epsilon$ is the interface thickness; $K \equiv c(c-1)(c-1/2) - \epsilon^2\nabla^2 c$, the chemical potential that characterizes the phase variation within the interface region; and $c$ is a phase function with $c = 1$ representing water and $c = 0$ representing oil[51] and (2) active stress exerted by extensile microtubule-based bundles in active fluid that was proportional to the orientational order of bundles, $\boldsymbol{\sigma}_a \equiv \alpha c \boldsymbol{D}$,[52] where $\alpha$ is an activity coefficient; $\boldsymbol{D} \equiv \int_{\boldsymbol{p}} \boldsymbol{pp}\psi d\boldsymbol{p}$, the local nematic order of bundles[53]; $\boldsymbol{p}$ represents the bundle orientation; and $\psi$ represents the probability distribution of bundles that satisfies the Smoluchowski equation[44,54]:

$$\frac{\partial \Psi}{\partial t} + \nabla \cdot (\dot{x}\Psi) + \nabla_p \cdot (\dot{p}\Psi) = 0, \tag{1}$$

where $x$ represents the center of mass of the bundle and $\nabla_p \equiv \partial(I - pp)/\partial p$ is the surface derivative on the unit sphere. These two forces were exerted on incompressible fluids ($\nabla \cdot u = 0$) to create flows ($u$) that satisfied the Navier–Stokes equation:

$$\rho\left(\frac{\partial u}{\partial t} + u \cdot \nabla u\right) = \nabla \cdot [-pI + \mu(\nabla u + \nabla u^T)] + F, \tag{2}$$

where $F \equiv T + \nabla \cdot \sigma_a$, the net body forces from interfacial tension and extensile bundles, $\rho$ is the fluid density, $p$ is the fluid pressure, and $\mu$ is the dynamic viscosity of fluids. The fluids were confined in no-slip boundaries whose geometries were identical to experimental containers and consisted of a circular side wall, a flat floor, and a curved ceiling (Fig. 1b). The boundary was filled with oil within which a compressed active droplet was immersed.

To solve the equations so as to determine the evolution of fluid flows, $u$, we initialized the flow field as quiescent fluids ($u = 0$) under uniform pressure ($p = 0$) with uniformly suspended isotropic bundles whose translational and orientational distributions were perturbed with 15 random modes (details of the random modes are provided in Supplementary Discussion S2).[44,55] Then we evolved the fluid flows for 3 hours with the finite element method based on the computational fluid dynamics software COMSOL Multiphysics[TM].[56,57] We made the assumption that the geometry of the interface remained invariant over time and water–oil interfacial fluctuation was negligible, based on our experimental results (Supplementary Video S1) and previous studies[32] showing that the water–oil interfacial tension is strong enough that the geometry of the droplet interface remains nearly unchanged over time. Thus, the phase function is independent of time:

$$c = H\left(1 - \frac{\sqrt{x^2 + y^2}}{r}\right), \tag{3}$$

where $H$ is the Heaviside step function. Details of the model, including the chosen values of parameters and explicit forms of each equation in three-dimensional components, are provided in Supplementary Discussion S2.

*Comparison between model prediction and experimental measurements.* To test the model's ability to describe our experimental system, we arranged two simulation systems with identical droplets ($r = 2.4$ mm, $h = 2$ mm) immersed in oil layers of different thicknesses ($\Delta = 1.1$ and $2.6$ mm). The simulation predicted that the droplet immersed in the thicker oil layer would develop a steady intradroplet circulatory flow, whereas the droplet immersed in the thinner oil layer would have chaotic flows (Figs. 6a&b inset). We then systematically varied the oil layer thickness ($\Delta = 0$–$9.6$ mm) in the model while maintaining the droplet geometry and analyzed the time-averaged COP within each droplet (Fig. 6b). The COP was sensitive to the oil layer thickness when the layer was thinner than ~2.2 mm, which suggests that the flows within and outside of the droplets were coupled. To reveal such coupling, we analyzed the flow profiles of azimuthal velocities across the water–oil interface (Fig. 6c), which showed that circulatory flows within the droplet induced a thin layer of circulatory flow in the oil with a layer thickness of 0.3 to 2 mm, whereas chaotic flows in the droplet did not induce net flows in the oil. However, analysis of the flow speed profiles revealed that oil near the interface developed flows with a thickness of ~1 mm (magenta curve in Fig. 6c inset), suggesting that flows within the active fluid droplet induce flows in the oil near the water–oil interface regardless of flowing state of the intradroplet active fluid (circulating or noncirculating).

*Characterization of active fluid–oil flow coupling with a cross-correlation function.* To gain deeper insight into coupling between the active fluid and the oil, we analyzed how the flows of active fluid near the interface (100 μm from the interface), $\boldsymbol{v}_w$, were correlated to oil flows, $\boldsymbol{v}_o$, by calculating the normalized same-time spatial cross-correlation function between $\boldsymbol{v}_w$ and $\boldsymbol{v}_o$:

$$\text{corr}(\Delta \boldsymbol{X}) \equiv \frac{\langle \boldsymbol{v}_o(\boldsymbol{x}+\Delta \boldsymbol{X}, t) \cdot \boldsymbol{v}_w(\boldsymbol{x}, t)\rangle_{x,t}}{\langle \boldsymbol{v}_o(\boldsymbol{x}, t) \cdot \boldsymbol{v}_w(\boldsymbol{x}, t)\rangle_{x,t}}, \tag{4}$$

where $\Delta \boldsymbol{X}$ represents the separation between a pair of active fluid and oil elements and $\langle\ \rangle_{x,t}$ indicates averaging over time in the active fluid region within 100 μm of the interface ($|r - x| \leq 100$ μm). To minimize the influence of the top and bottom boundaries[12,58-60] in our correlation analysis, we only considered the flows at the midplane ($z = h/2$). To reveal the coupling range between flows of active fluid and oil, we averaged the correlation function over the orientation:

$$C(\Delta X) \equiv \langle \text{corr}(\Delta \boldsymbol{X})\rangle_{|\Delta \boldsymbol{X}|=\Delta X}. \tag{5}$$

Our analysis revealed that the correlation function decayed nearly exponentially with increasing distance between the active fluid element and the oil element ($C \sim e^{-\Delta X/L}$, where $L$ is the correlation length), which suggests that the active fluid–oil interaction is short ranged (Fig. 7a). To quantify the interaction range, we extracted the correlation length, $L$, and then analyzed the correlation length as a function of oil layer thickness (Fig. 7b). Our analysis revealed that the correlation length linearly increased with the oil layer thickness ($L \approx \Delta$), eventually reaching saturation ($L \approx 1.4$ mm). The linear increase indicated that the active fluid flows were coupled to the oil flows throughout the oil region. Hence, the geometry of the oil (such as oil layer thickness) affected the active fluid flows, which is consistent with our observation that the COP in the active fluid changed rapidly with the oil layer thickness when the thickness was small ($\Delta <$ 2.2 mm in Fig. 6b). The saturation indicated that the range of active fluid–oil interaction has an upper limit (~1.4 mm) above which the motion of oil elements did not affect the active fluid. This is consistent with our observation that the COP in the active fluid was independent of the oil layer thickness when the oil layer was sufficiently thick ($\Delta > 2.2$ mm in Fig. 6b). Moreover, the scale of the analyzed correlation length (~1.4 mm; Fig. 7b) was consistent with the observed coupling length (~1 mm; Fig. 6c inset). This consistency, along with simulated shear stress analysis (Supplementary Discussion S3), supports the assertion that active fluid and oil interact across the water–oil interface with a millimeter–scale interaction range. Further analyses of the role of interfacial properties, such as viscosity contrast and interfacial tension, on the simulated results are provided in Supplementary Discussion S4. Overall, the simulation outcomes were qualitatively consistent with our experimental characterizations of the coupling of intradroplet and extradroplet flows (Figs. 3&5). This consistency demonstrates that the model can describe active fluid–oil coupling and how the coupling influences the flows inside the active fluid droplet.

*Limitations of the model.* The simulation failed to match two of our experimental outcomes: First, the simulation showed that the intradroplet fluid flows became chaotic when oil layers were thicker than ~5.6 mm (Fig. 6b), whereas in our experimental data, the intradroplet fluid flows were insensitive to the oil arrangement when the oil layers were thicker than ~4 mm (Fig. 3a–d). Second, the simulation predicted the induction of circulatory flows in oil driven by intradroplet circulation (Fig. 6c), whereas in the experiments, oil did not develop net flows regardless of how the fluid flowed within the droplets (Fig. 5c). It is possible that these discrepancies could be mitigated by allowing the interface to deform following the rules of spontaneous phase separation of oil and water (the Cahn-Hilliard model).[61]

**Intradroplet circulatory flows were triggered and suppressed in real time with novel millifluidic devices.** Our experimental results show that the formation of intradroplet circulatory flows depends on droplet geometry (Fig. 2) and oil layer thickness (Fig. 3). This suggests that millifluidic devices could control the formation and suppression of circulatory flows in real time by changing the droplet shape or oil layer thickness. We demonstrated this by designing and testing two such devices.

*Device that changes oil layer thickness with a movable wall.* To manually tune the oil layer thickness, we developed a cylindrical container that compressed an active droplet ($h = 2$ mm, $r \approx 2.4$ mm) and had one movable wall (pink blade in Figs. 8a&b) that could be moved toward the droplet to reduce the thickness of the oil layer adjacent to one part of the droplet (minimum oil layer thickness $\Delta_m \approx 1.2$ mm; Fig. 8b left). When the wall was near the droplet, the active fluid flowed chaotically (|COP| ≲ 0.2; Fig. 8c). When we increased the oil thickness by moving the wall away from the droplet ($\Delta_m \approx 2.6$ mm; Fig. 8b middle), the active fluid developed circulatory flows in ~30 minutes (COP = 0.4–0.6; Fig. 8c). The circulatory flows lasted for ~1 hour and then transitioned to chaotic flows after we moved the wall back toward the droplet ($\Delta_m \approx 1.2$ mm; Fig. 8b right, Supplementary Video S3). These results show that it is possible to develop and inhibit intradroplet circulatory flows locally in real time without physically contacting the droplet.

*Device that deforms droplet with a movable ceiling.* To manually shape the droplet, we compressed the droplet in a cylindrical container with a movable ceiling (Fig. 8d). We first compressed the droplet to a short cylinder-like shape ($r \approx 2.0$ mm, $h = 2$ mm), which supported the formation of circulatory flows (COP = 0.4–0.6, Fig. 8e). The circulatory flow persisted for ~40 minutes before it was manually suppressed (|COP| ≲ 0.2) by lifting the ceiling ($h = 3$ mm), which shaped the droplet into a taller cylinder-like shape ($r \approx 1.7$ mm, $h = 3$ mm). Conversely, intradroplet circulatory flow could be manually triggered by deforming the droplet from a taller ($r \approx 1.7$ mm, $h = 3$ mm) to a shorter ($r \approx 2.0$ mm, $h = 2$ mm) cylinder-like shape (Fig. 8f, Supplementary Video S4). These results demonstrate that manually shaping the droplet can turn intradroplet circulatory flows on and off locally. These findings pave the way for designing fluidic devices that can shape deformable boundaries to direct the self-organization of confined active fluids in real time.

**Limitations of the study.** This study focused on microtubule-based active fluid and thus the results may not be generalizable to other active fluids. Another limitation of this study is that both the models and experiments neglected interfacial fluctuation, because interfacial tension in this system was strong enough to inhibit interfacial fluctuation. Low interfacial tension can distort interfaces[62,63] which could alter flow coupling across the interface. Future research could investigate flow coupling in systems with different interfacial tensions by incorporating vesicles[62] or varying interfacial surfactant concentrations.[64]

## Conclusions
This work demonstrates that the self-organization of a water-in-oil droplet of microtubule-based active fluid is influenced by flow coupling across the water–oil interface with a millimeter–scale coupling length. Our experimental data and simulation results indicate that active fluid within the droplet can induce flows in the oil within this coupling range and that disturbances outside of the droplet, such as reducing the thickness of the oil layer to be less than the coupling range, can influence the microtubule network structure and active stress distribution inside the droplet and thus impact the intradroplet flows. While previous studies have shown that the formation of circulatory flows depends on the confinement boundary geometry,[11,13-15,18] this work is the first to highlight the role of boundary conditions on the formation of circulatory flows—specifically the role of fluid boundaries and hydrodynamic coupling across active–passive boundaries in confined active fluid systems.

We also developed two millifluidic devices that can trigger and suppress intradroplet circulatory flows in real time: one suppresses intradroplet circulatory flows without contacting the droplet by manipulating the oil layer thickness and thus disturbing the active fluid–oil coupling and the other compresses droplets to desired height and radius combinations that support or suppress intradroplet circulatory flows. These novel devices provide the biology community with *in vitro* model systems to probe how the deformation of cell membranes or the disturbance of fluids around cells influences intracellular activities. These approaches could also be used in the development of treatment modalities for cells that are sensitive to biomedical approaches and can only be treated with physical methods, such as deforming cells and imposing shear flows.[65,66] In mechanical engineering, these systems pave the way for designing machines driven by active fluid with adjustable power output.[67]

**Methods**

**Fabricate a millifluidic device to confine active water-in-oil droplets.** To confine the active fluid in a cylinder-like water-in-oil droplet, we designed a millifluidic device that compressed the droplet between a pair of plates separated by a height ($h$) of 1 to 2 mm. Because the compressed droplet was self-propelling,[26,32,40] we immobilized the droplet by curving the upper plate surface into a half-oblate spheroidal dome with a shallow cylindrical well (height 0.2 mm; radius 1 mm) at the dome center (Supplementary Fig. S1a). The curved surface and shallow well immobilized the droplet without significantly impacting the intradroplet active fluid behaviors (Supplementary Discussion S1). Finally, to load the oil and active fluid into the millifluidic device, we drilled a 3.6-mm-long loading channel 2 mm wide and 1.7 mm high, and joined the channel to a 2.4-mm-long neck that had the same width (2 mm) but shorter height (1.4 mm). To fabricate the device, we sketched the 3D design in SolidWorks, programmed the corresponding tool paths in Esprit, and used the tool paths to end-mill a $610 \times 38 \times 6.4$ mm$^3$ acrylic rectangular bar with computer numerical control (McMaster 1227T222). The milled chip was then cleaned with sequential 10-minute sonication in detergent (Sigma-Aldrich Z805939), ethanol, and 100 mM potassium hydroxide solution and glued to a fluorophilically-treated glass slide (VWR GWBJ17) with epoxy (Bob Smith Industries BSI-201) to complete the device fabrication.[68]

**Prepare the compressed water-in-oil droplet.** We prepared microtubule–kinesin active fluid according to our previous protocols[32,69] and pipetted the active fluid through the loading channel into the chamber of the fabricated millifluidic device that was filled with oil (hydrofluoroether, 3M Novec 7500; Supplementary Fig. S1a). To prevent the microtubule and kinesin proteins from contacting the oil (and thus denaturing), we doped the oil with 1.8% surfactant (perfluoropolyether–polyethylene glycol–perfluoropolyether, RAN Biotechnologies 008-FluoroSurfactant)[68] to stabilize the protein near the water–oil interface. (Previous studies of systems with microtubule-based active fluid interfacing with oil showed that microtubules could be centrifuged to the water–oil interface and would then form a two-dimensional active nematic layer.[18,30,32,62] In our system, we did not centrifuge the samples, and though a small portion of microtubules were spontaneously deposited onto the interface and formed 2D active nematics [Supplementary Discussion S5], the majority of microtubules remained in the bulk and induced active fluid flows [Supplementary Video S1].) The active fluid loaded into the channel then formed a water-in-oil droplet that was compressed between the ceiling and floor of the chamber. The compression deformed the droplet into a cylinder-like shape[39] whose height ($h$) depended on the ceiling–floor separation of the chamber and whose radius ($r$) depended on the pipetted fluid volume. After injecting the active droplet, we sealed the channel with epoxy. However, after the channel was sealed, air bubbles sometimes formed in the sample and affect the experimental outcomes. To keep the bubbles away from the droplet, we tilted the sample to direct the air bubbles to exit the chamber into the loading channel through the neck. The neck's smaller opening prevented the bubbles from reentering the chamber (Supplementary Fig. S1a close-up).

In our experiments, we first varied the droplet radius in millifluidic devices with chambers of radii $R = 3.5$ and 5 mm (Fig. 2) to examine the role of the droplet radius in the formation of the intradroplet circulatory flows. Each of the devices had a half-spheroidal dome with a vertical semi-axis of $R_a = 0.25$ mm and a horizontal semi-axis matching the chamber radius, $R_b = R$ (Supplementary Fig. S1a). We then immersed the droplets in oil layers of various thicknesses (Fig. 3). The oil was contained in the chamber whose radius was the sum of the droplet radius and oil layer thickness $R = r + \Delta$ (Fig. 1b). To minimize the influence of the ceiling shape on the experimental outcomes, we chose a fixed half-spheroidal ceiling ($R_a = 0.5$ mm, $R_b = 5$ mm) to cover the chamber. When the chamber was smaller than the ceiling ($2R < 2R_b$), the ceiling was trimmed to fit into the chamber; conversely, when the chamber was larger than the ceiling ($2R > 2R_b$), the ceiling was extended horizontally to match the chamber size (Supplementary Fig. S1b&c).

**Image and analyze flows.** To observe the flows of the active fluid and oil, we doped the active fluid with 0.0004% Alexa 488-labeled 3-μm tracer particles (Polysciences 18861-1), which could be imaged with a green fluorescent protein filter cube (excitation: 440–466 nm, emission: 525–550 nm, Semrock 96372). To characterize the flow behaviors, we imaged the tracers for 3 hours and tracked their trajectories $\boldsymbol{x}_i(t)$ with the Lagrangian algorithm.[70] The trajectories revealed the evolution of the particles' velocities, $\boldsymbol{v}_i(t) \equiv d\boldsymbol{x}_i(t)/dt$, and allowed us to calculate the COP, $\text{COP}(t) \equiv \langle v_{i,\theta}(t)/|\boldsymbol{v}_i(t)|\rangle_i$, where $v_{i,\theta}$ is the azimuthal component of velocity $\boldsymbol{v}_i$ of the $i^{\text{th}}$ tracer and $\langle\ \rangle_i$ indicates averaging of the tracer particles (Fig. 1d). To quantify the coherence of the intradroplet circulatory flows, we measured the time-averaged COP: $\langle \text{COP}(t)\rangle_t$ (Figs. 2&3a–d). To reveal flow rates of intradroplet circulatory flows, we measured the flow profiles of the azimuthal velocities across the droplet, $\langle v_\theta(d)\rangle \equiv \langle v_{i,\theta}(d,t)\rangle_{t,i}$, where $d$ is the distance from the water–oil interface (Figs. 3e–h). To reveal the structure of the flows in the droplet, we measured the time-averaged, normalized velocity fields, $\boldsymbol{V}(\boldsymbol{x}) \equiv \langle\frac{\boldsymbol{v}(\boldsymbol{x},t)}{\langle|\boldsymbol{v}(\boldsymbol{x},t)|\rangle_x}\rangle_t$, and vorticity distributions, $\Omega(\boldsymbol{x}) \equiv \langle\frac{\omega(\boldsymbol{x},t)}{3\sigma(t)}\rangle_t$, where $\boldsymbol{v}(\boldsymbol{x},t)$ is the velocity field calculated from the sequential tracer images with the particle image velocimetry algorithm[71]; $\omega(\boldsymbol{x},t) \equiv [\nabla \times \boldsymbol{v}(\boldsymbol{x},t)]_z$, the corresponding vorticity distribution; and $\sigma(t) \equiv \text{std}[\omega(\boldsymbol{x},t)]$, the standard deviation of the vorticity (Fig. 1c).[11]

To characterize flows in oil, we doped the oil with 0.002% Alexa 488-labeled 1-μm tracer particles (Polysciences 18860-1) and conducted the same observations and analyses for active fluid flows (Figs. 5b&c). To reveal flow activities near water–oil interfaces, we measured the flow speed profiles $\langle|\boldsymbol{v}(\rho)|\rangle \equiv \langle|\boldsymbol{v}_i(\rho,t)|\rangle_{t,i}$, where $\rho$ is the radial coordinate relative to the water–oil interface (Fig. 5c&d).

**Image and analyze microtubule network structure.** To characterize the influence of the oil layer thickness on the intradroplet microtubule network structure, we imaged the microtubules at the droplet midplane with confocal microscopy (Leica SP5 point scanning confocal microscope). The microtubules were labeled with Alexa 647 (according to our previous protocol[69]), excited with a 633-nm helium-neon laser, and observed in a 633–647 nm window. To reveal the microtubule arrangement in a circulating active fluid, we used our data to select the droplet shape ($r \approx 2.4$ mm, $h = 1$ mm) and oil layer thickness ($\Delta \approx 2.4$ mm) that supported the formation of intradroplet circulatory flows (Fig. 3b). To observe the time-averaged network structure, we imaged the network every 2 seconds for 1 hour and then analyzed the images with the snake algorithm to extract the network structure, which consisted of unit-length segments (Fig. 4b&c).[41] We stacked the segment orientations from each image and then counted these orientations to reveal the orientational distribution of the microtubule bundles (Fig. 4d). The microtubule bundle orientations were measured near the water–oil interface, ~400 μm from the interface, and at the droplet center (Fig. 4a) and the measurements were repeated on another sample with the same droplet shape but different oil layer thickness ($\Delta \approx 1.1$ mm) where intradroplet circulation was suppressed.

**Analyze microtubule network dynamics.** To further characterize the influence of oil layer thickness on the dynamics of the microtubule network, we first measured time-averaged velocity fields and vorticity maps by analyzing the sequential confocal images of the microtubule network near water–oil interfaces with the particle image velocimetry algorithm (Figs. 4e&h).[71] To reveal the corresponding bundle alignments in these flows, we measured the time-averaged director fields $\hat{n}$ and nematic order parameter maps (Figs. 4f&i) by first analyzing the bundle orientational tensor, $\boldsymbol{D} \equiv \langle \boldsymbol{pp} \rangle_t$, where $\boldsymbol{p}$ represents the extracted bundle orientation from the confocal images (Fig. 4c) and $\langle \ \rangle_t$ represents averaging over time. Then, we determined the nematic order tensor, $\boldsymbol{Q} \equiv \boldsymbol{D} - \boldsymbol{I}/2$, calculated the maximum eigenvalue, $\lambda_m$, and determined the nematic order parameter as NOP $= 2\lambda_m$ and the director $\hat{n}$ as the corresponding eigenvector.[53] To show the dynamics that resulted from these bundle configurations, we measured the time-averaged active stress maps (Fig. 4g&j) by calculating the magnitude of the bundle orientational tensor, $\sqrt{\boldsymbol{D}:\boldsymbol{D}}$. $\sqrt{\boldsymbol{D}:\boldsymbol{D}}$ represents active stress because, according to Gao *et al.*'s model,[26] active stress tensor is defined as $\boldsymbol{\sigma}_a \equiv \alpha \boldsymbol{D}$ where $\alpha$ is an activity coefficient, which is a constant in our active fluid, and active stress magnitude is determined as $\sqrt{\boldsymbol{\sigma}_a:\boldsymbol{\sigma}_a} = \alpha\sqrt{\boldsymbol{D}:\boldsymbol{D}}$, which is proportional to $\sqrt{\boldsymbol{D}:\boldsymbol{D}}$.

**Design fluidic device with movable wall.** To change the thickness of oil surrounding a droplet in real time, we designed a fluidic device with a movable wall, inspired by the mechanical iris that is used to adjust the aperture size of optical devices.[72] However, a conventional mechanical iris consists of at least six blades, and fabricating a six-blade mechanical iris at the micron scale was challenging because at this scale the blades bent spontaneously in our fabrication process and could not be assembled. Therefore, we simplified the design to contain only one blade that was thick enough (1 mm) to remain flat (pink components in Supplementary Fig. S8). The single blade was pinned to a cylindrical container where the pin was constrained in a radial groove in the container and would move toward the chamber center when the blade was rotated counterclockwise. This blade rotation caused the midpoint of the blade to move 1.4 mm closer to the chamber center, thus shrinking the size of the chamber (Supplementary Fig. S8, right). Conversely, when the blade was rotated clockwise, the chamber expanded (Supplementary Fig. S8, left). We enclosed the container by gluing fluorophilic glass to the bottom of the chamber and a curved ceiling to the top of the container. Once the glue was cured, oil and active fluid were pipetted into the chamber via the loading channel and the channel was then sealed with epoxy. While the channel was sealed, the chamber remained semi-open because the movable blade required a gap between the ceiling and floor. This gap would allow oil to evaporate and create air bubbles that could influence experimental outcomes. To minimize the influence of oil evaporation, we overflowed the gap and blade with oil so that the blade was below the oil surface. This arrangement allowed us to rotate the blade without exposing the chamber to air.

**Design fluidic device with movable ceiling.** To control the shape of the droplet in real time, we designed a millifluidic device whose ceiling could be tuned manually. To fabricate the device, we separated the ceiling from the rest of the device and attached a handle to manually move the ceiling vertically. We also attached a hanger to hold the ceiling on top of the device chamber (Supplementary Fig. S9a). To control the vertical position of the ceiling, we designed three platforms with different heights that were placed outside the edge of the device chamber where we could hang the ceiling to adjust its height from $h = 1$ to 3 mm (Supplementary Fig. S9b). To minimize the influence of oil evaporation, we immersed the system (including the device chamber and ceiling) in an oil bath enclosed in a Petri dish.

**Data availability:** The data that support the findings of this study are available from the corresponding author upon reasonable request.

## Acknowledgements

We thank Dr. Zvonimir Dogic for the gift of K401-BCCP-H6 plasmids which were used for expressing the kinesin motor proteins for driving the microtubule-based active fluid. We thank Drs. Yuan-Nan Young and David B. Stein for the insightful discussion on interpreting our experimental data. We thank Victoria M. Bicchieri for her assistance on collecting confocal data with the Leica SP5 point scanning confocal microscope in the Life Sciences and Bioengineering Center at Worcester Polytechnic Institute. We thank Ellie Lin from Lin Life Science for her assistance on editing the manuscript to enhance its flow and readability.

K.-T.W. acknowledges support from the National Science Foundation (NSF-CBET-2045621). This research was performed with computational resources supported by the Academic & Research Computing Group at Worcester Polytechnic Institute. We acknowledge the Royal Society of Chemistry for adapting the figure from Bate *et al.* on Soft Matter.[36] We acknowledge the Brandeis Materials Research Science and Engineering Center (NSF-MRSEC-DMR-2011486) for use of the Biological Materials Facility. C.-C.C. acknowledges support for the numerical studies from the Headquarters of University Advancement at the National Cheng Kung University, sponsored by the Ministry of Education, Taiwan.

## Author Contributions

Y.-C.C., B.J., C.-C.C., and K.-T.W. performed the research and designed the experiments; Y.-C.C. fabricated the millifluidic devices, prepared the samples, tracked the flows, and reconstructed the microtubule network structures; Y.-C.C., B.J., and C.-C.C. established the continuum simulation platform on modeling active droplet systems; Y.-C.C. and K.-T.W. organized and analyzed the data; Y.-C.C., C.-C.C., and K.-T.W. wrote the manuscript; and K.-T.W. supervised the research. All authors reviewed the manuscript.

## Additional Information

**Competing interests statement.** The authors declare that they have no competing interests.

**Correspondence.** Correspondence and requests for materials should be addressed to K.-T.W. (kwu@wpi.edu). Active fluid simulation questions should be addressed to C.-C.C. (chuehcc@mail.ncku.edu.tw).

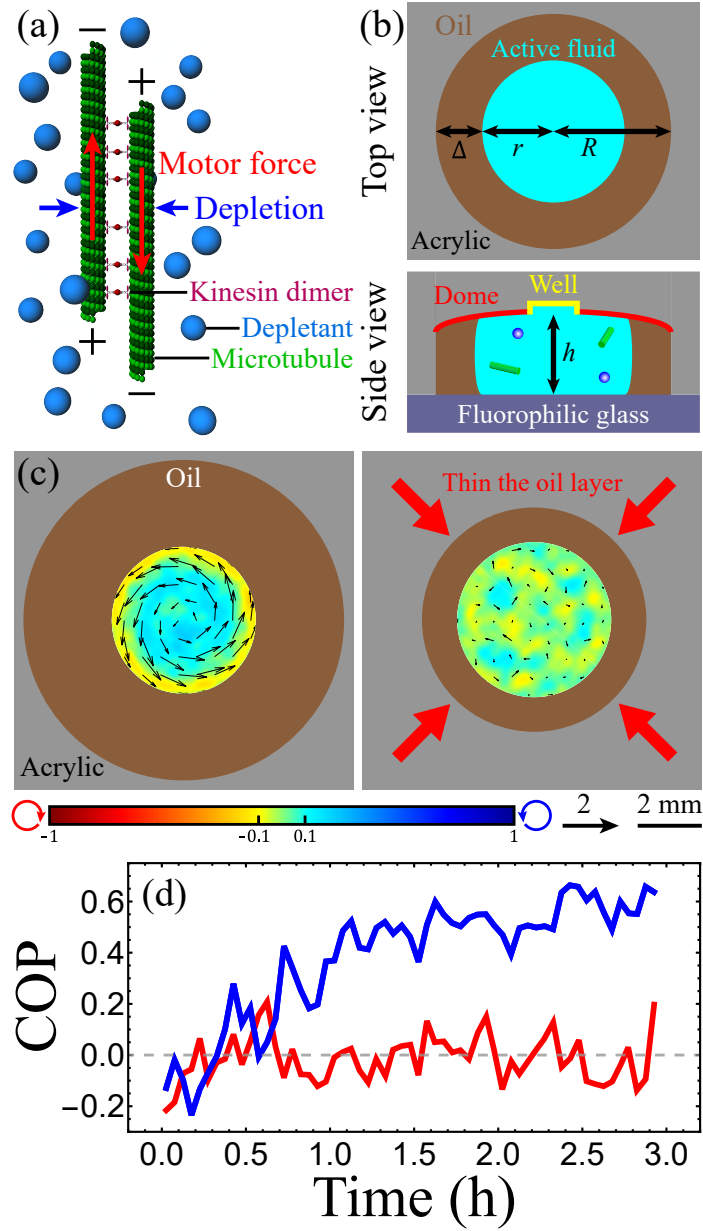

Fig. 1. Flows within an active droplet immersed in oil are affected by the thickness of the oil layer. (a) Mechanism behind the activity of microtubule–kinesin active fluid. Microtubules are bundled by depletion (blue arrows) and bridged by kinesin motor dimers that slide pairs of antiparallel microtubules apart (red arrows). Figure adapted from Bate et al.[36] (b) Schematic of fluidic device used to compress droplets. The device ceiling was curved into a dome-like shape (red), and a well was drilled (radius 1 mm, depth 0.2 mm) on top of the dome (yellow). The curved ceiling caused the active droplet to remain stationary at the dome center. (c) An active droplet ($r \approx 2.4$ mm, $h \approx 2$ mm) confined in an oil layer whose thickness was 2.6 mm developed a circulatory flow (left). The circulatory flow was suppressed by decreasing the oil layer thickness to 1.1 mm (right). Black arrows represent time-averaged velocity flow fields normalized by the flow mean speed. The color maps represent normalized vorticity distributions, with red and blue representing clockwise and counterclockwise vorticities, respectively. Videos of the flows in both systems are available in Supplementary Video S2. (d) Evolution of circulation order parameter (COP) of flows within the active droplets in c (blue: 2.6-mm oil thickness; red: 1.1-mm oil thickness).

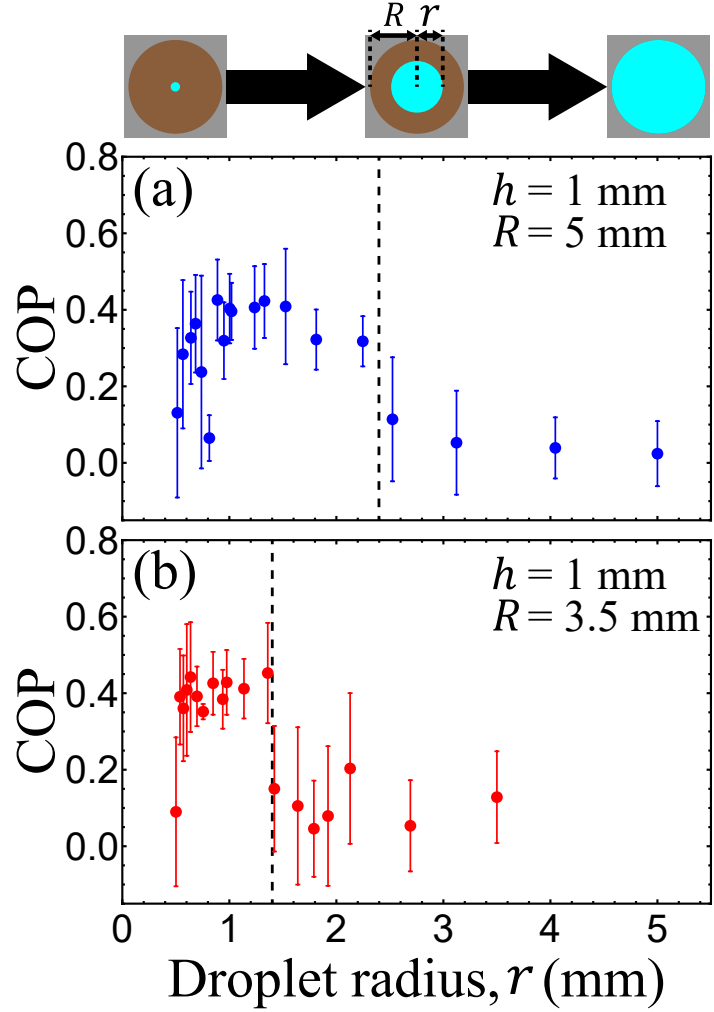

Fig. 2. The intradroplet circulatory flows depended on the dimensions of the droplets and oil. (a) Droplets immersed in an oil of radius $R = 5$ mm developed circulatory flows when their radii were $r \lesssim 2.4$ mm (black dashed line). Enlarging the droplets above this limit suppressed the formation of circulatory flows. (b) However, such critical radius was altered when the oil was shrunken to a radius of $R = 3.5$ mm. In this case, the circulation transition was shifted to $r \approx 1.4$ mm (black dashed line). The shift implies that the intradroplet circulatory flows were affected by the oil dimensions and by the droplet radii. Each error bar represents the standard deviation of the time-averaged circulation order parameter (COP).

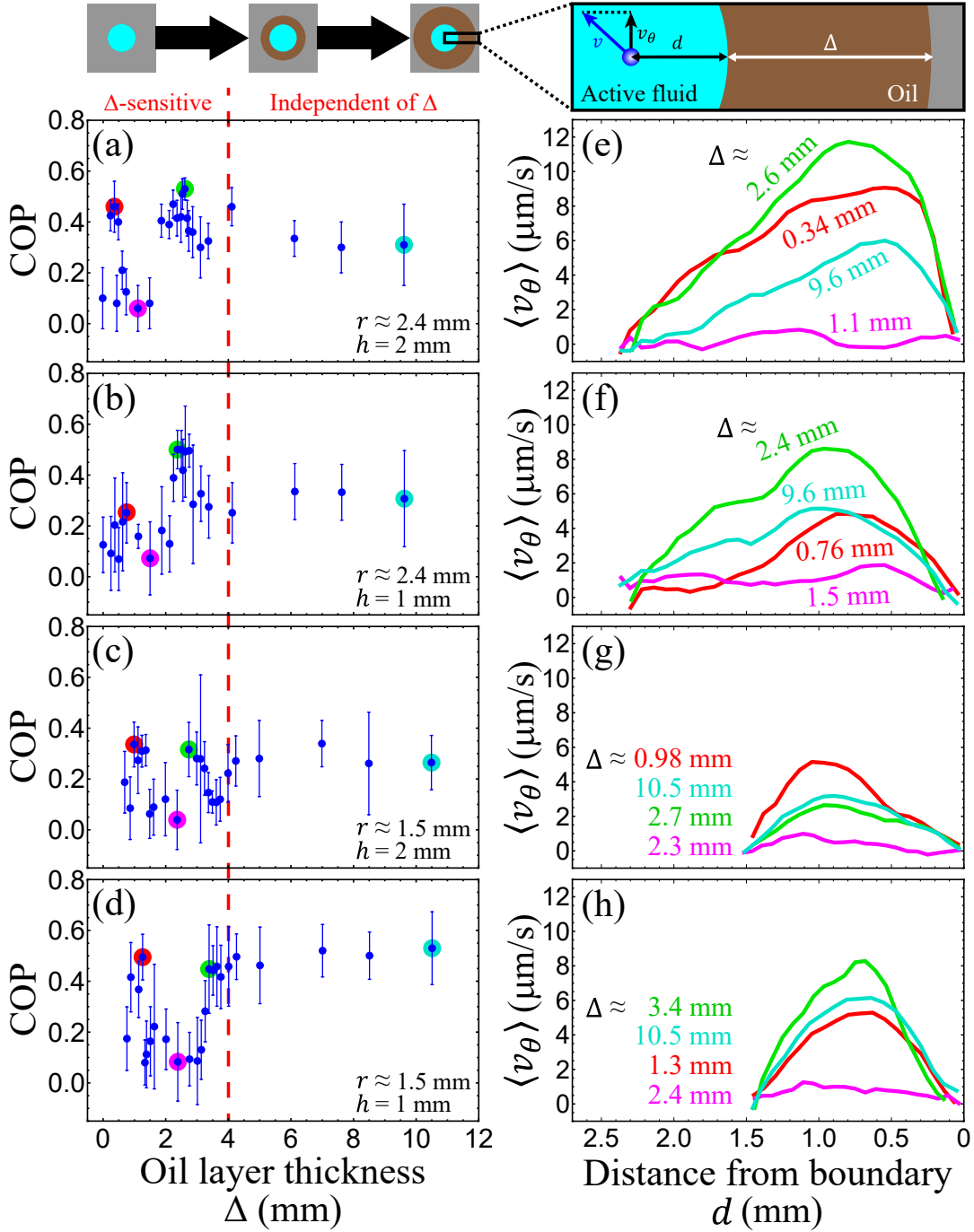

Fig. 3. Intradroplet circulatory flow depended on the thickness of the surrounding oil layer, $\Delta$. (a–d) The circulation order parameter (COP) as a function of $\Delta$ for four droplet shapes. Across these droplets, the intradroplet flows varied rapidly when the oil layer was thinner than ~4 mm (red dashed line), whereas oil layers thicker than this limit did not affect intradroplet flows. Each error bar represents the standard deviation of the time-averaged COP. (e–h) Flow profiles of azimuthal velocity, $v_\theta$, taken at droplet midplane. Each curve represents the averaged flow velocity of the data point shaded in the same color in the same row of a–d. These curves indicate that varying the thickness of the oil layer altered the net flow rates of the intradroplet flows.

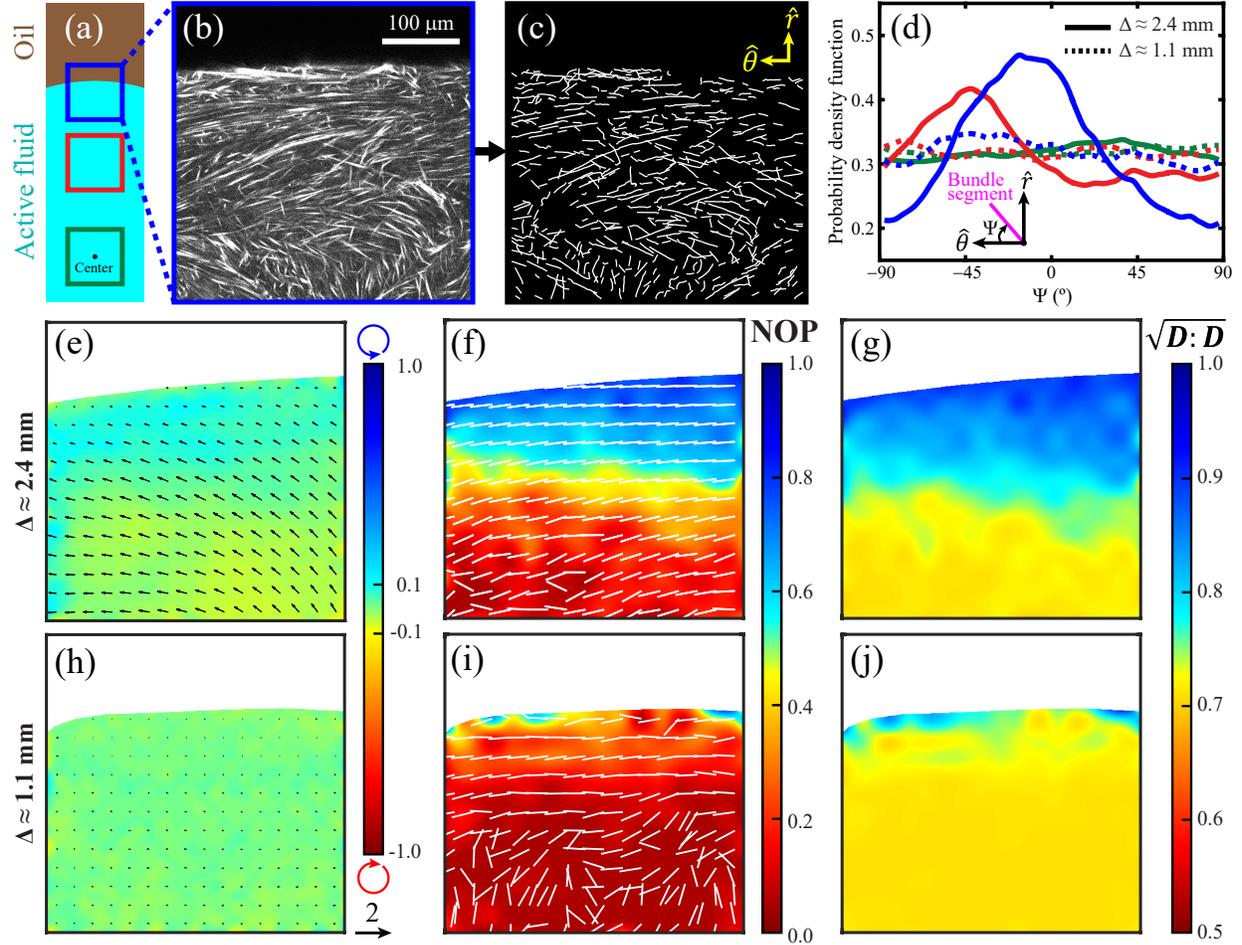

Fig. 4. Intradroplet circulatory flows were accompanied by the formation of a weak nematic layer in the microtubule network structure near the water–oil interface. A droplet (radius = 2.4 mm; height = 1 mm) was immersed in an oil layer of thickness ($\Delta$) 2.4 mm and developed an intradroplet circulatory flow. (a) Microtubule network structures were imaged with a confocal microscope at three locations: the water–oil interface (blue), ~400 μm from the interface (red), and the droplet center (green). (b) Confocal image of the microtubule network at the droplet midplane. (c) Corresponding network structure extracted with the snake algorithm.[41] (d) Orientational distributions of microtubule bundles when intradroplet flows were circulating (solid, $\Delta \approx 2.4$ mm) and not circulating (dashed, $\Delta \approx 1.1$ mm). Blue, red, and green curves represent the locations within the droplet indicated in panel a. In circulating flows, most microtubule bundles near the interface were aligned at angles of ~15° from the interface (solid blue curve), and most microtubule bundles 400 μm from the interface were aligned at angles of ~45° (solid red curve). In noncirculating flows, the bundles were oriented isotropically (dotted curves). (e–j) Time-averaged velocity fields and vorticity maps of microtubule flows (panels e&h, plotted as in Fig. 1c), director fields and nematic order parameter (NOP) maps of microtubule bundles (panels f&i), and $\sqrt{D:D}$ maps of microtubule network (panels g&j) near the water–oil interface. $D$ is bundle orientational tensor defined as $D \equiv \langle pp \rangle$ where $p$ represents the orientation of a bundle segment. In Gao et al.'s model,[26] $\sqrt{D:D}$ is proportional to magnitude of active stress $\sigma_a$ so $\sqrt{D:D}$ maps can be interpreted as active stress maps. The first row (panels e–g) represents a droplet immersed in a 2.4-mm-thick oil layer, and the second row (panels h–j) represents a droplet immersed in a 1.1-mm-thick oil layer.

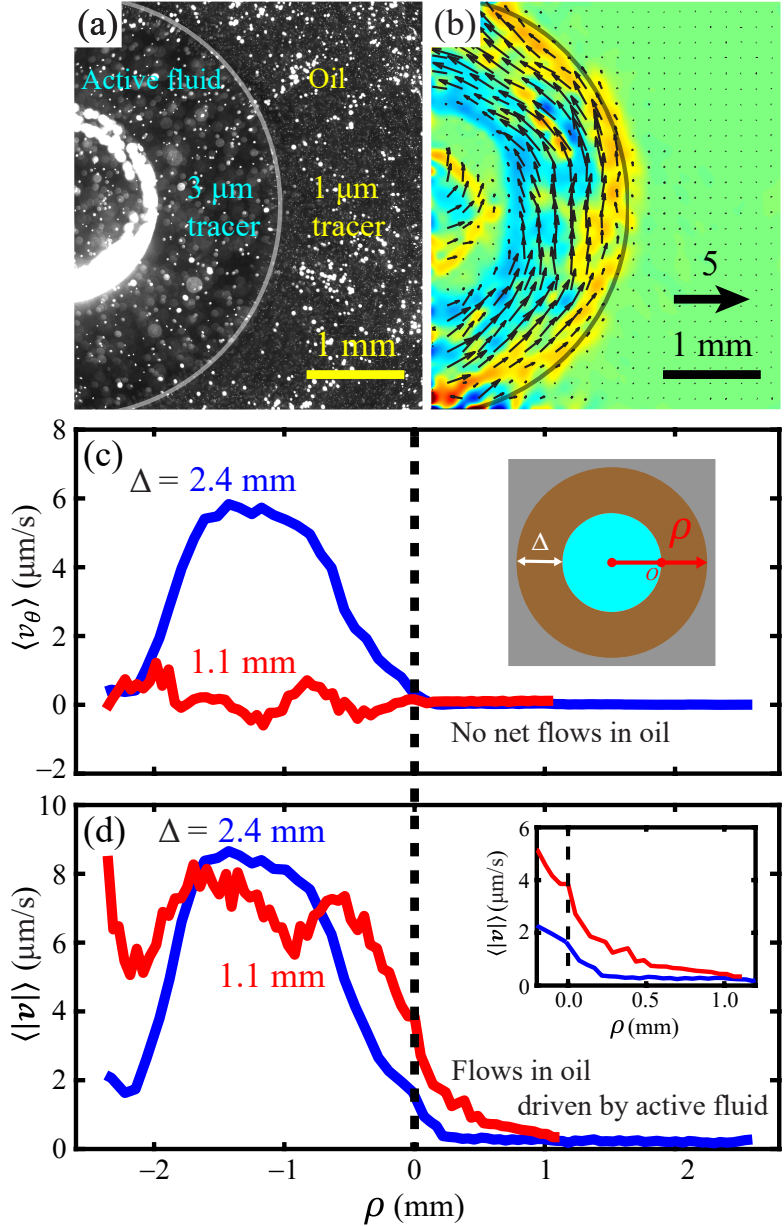

Fig. 5. Active fluid droplets induced chaotic flows in the surrounding oil. (a) Micrograph of an active droplet ($r \approx 2.4$ mm, $h = 1$ mm) immersed in oil ($\Delta \approx 2.4$ mm). To track the flows, both the active fluid and oil were doped with tracers. The gray curve represents the water–oil interface. (b) Time-averaged normalized velocity field and vorticity map revealed circulatory flow in the droplet but no net flow in the oil. The velocity field and vorticity map were plotted as in Fig. 1c. (c) Midplane flow profiles of azimuthal velocities of two droplets with the same shape ($r \approx 2.4$ mm, $h = 1$ mm) but different oil layer thicknesses ($\Delta \approx 2.4$ mm, blue curve; $\Delta \approx 1.1$ mm, red curve) show that the thicker oil layer supported intradroplet circulatory flows, whereas the thinner layer did not. The dashed vertical line indicates the water–oil interface. Regardless of how the active fluids flowed within the droplets, the oil developed no net flow. Inset: Schematic of horizontal axis of the plot, $\rho$. $\rho > 0$ represents the oil region; $\rho < 0$ represents the droplet region. (d) The flow speed profiles for both droplets extended across the water–oil interface into the oil region ($\rho > 0$). This shows that the active fluid flows near the interface drove flow in the nearby oil. Inset: Close-up of the profiles near the water–oil interface.
19

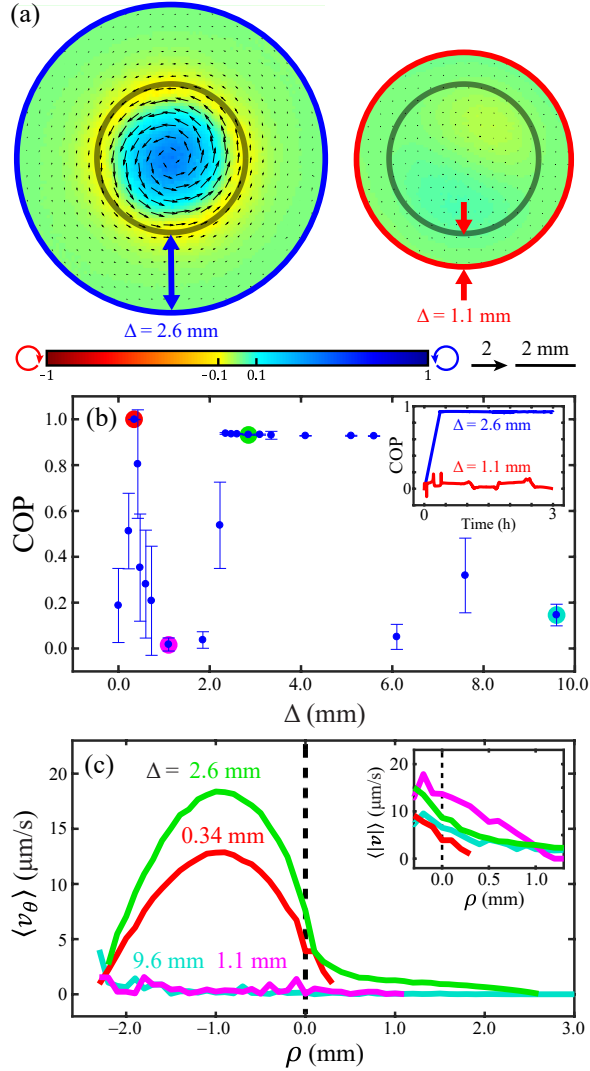

Fig. 6. Simulation of behavior of compressed water-in-oil droplets based on continuum active fluid equations. (a) Simulated time-averaged velocity fields and vorticity maps from the midplanes of compressed droplets ($r = 2.4$ mm, $h = 2$ mm) immersed in oil layers of different thicknesses ($\Delta$). The velocities and vorticities were plotted as in Fig. 1c. Gray curves represent the water–oil interfaces; blue and red curves represent the no-slip boundaries. Droplets immersed in a thicker oil layer ($\Delta = 2.6$ mm) developed circulatory flows (left), whereas in those with a thinner oil layer ($\Delta = 1.1$ mm) the circulatory flows were suppressed (right). Simulated instantaneous velocity fields and vorticity maps in $xy$-, $yz$-, and $xz$-midplane cross-section for various oil layer thicknesses are available in Supplementary Fig. S6. (b) Circulation order parameter (COP) as a function of the layer thickness of the oil that surrounds a compressed droplet ($r = 2.4$ mm, $h = 2$ mm). The intradroplet flows were sensitive to oil layer thickness when the thicknesses were thinner than ~2.2 mm. Each error bar represents the standard deviation of the time-averaged COP. Inset: Evolution of COPs within the droplets described in panel a. (c) Flow profiles of azimuthal velocities, $v_\theta$, taken at droplet midplanes. The horizontal axis represents the radial axis in cylindrical coordinate with the origin shifted to the droplet interface (Fig. 5c inset). Each curve represents the averaged flow velocity of the data points shaded in the same color in b. Dashed lines represent the water–oil interface. These flow profiles demonstrated that intradroplet circulatory flows induced a thin layer (0.3–2 mm) of circulatory flow in the oil near the interface, indicating that flows within and outside the interface were coupled. Inset: Flow speed profiles near the droplet interface.

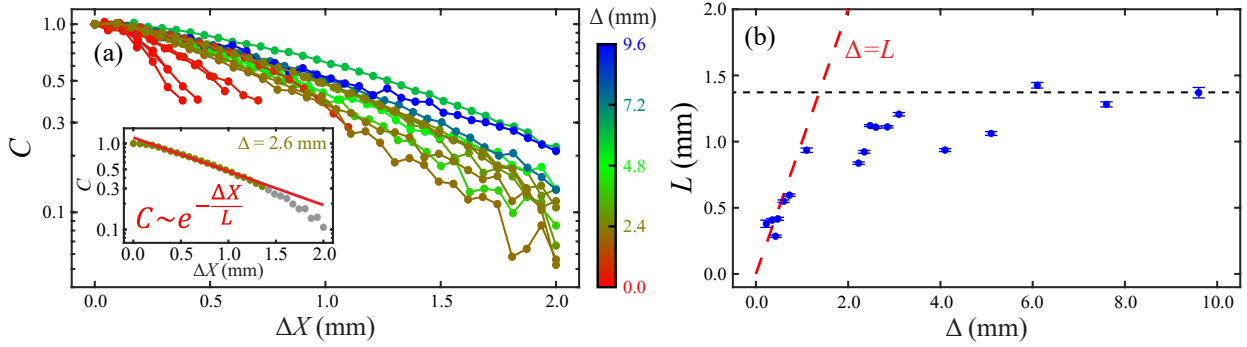

Fig. 7: Simulations revealed that flows inside and outside of water-in-oil droplets ($r = 2.4$ mm, $h = 2$ mm) were correlated with a correlation length that increased with the oil layer thickness. (a) Normalized same-time velocity–velocity spatial cross-correlation functions between active fluid near the water–oil interface (within 100 μm) and oil for various oil layer thicknesses ($\Delta$). The correlation functions decayed with increasing distance between the active fluid element and the oil element ($\Delta X$). Inset: The initial decay length (defined as the correlation length) $L$ of the correlation function was extracted by excluding the correlation data below 0.3 (gray dots) and fitting the remaining correlation data (olive dots) to an exponential function: $e^{-\Delta X/L}$ with $L$ as a fitting parameter (red line). (b) The correlation length increased linearly with the oil layer thickness (dashed red line) before reaching its maximum (black dashed line). Error bars represent uncertainties in fitting the correlation function to an exponential function (inset in a).

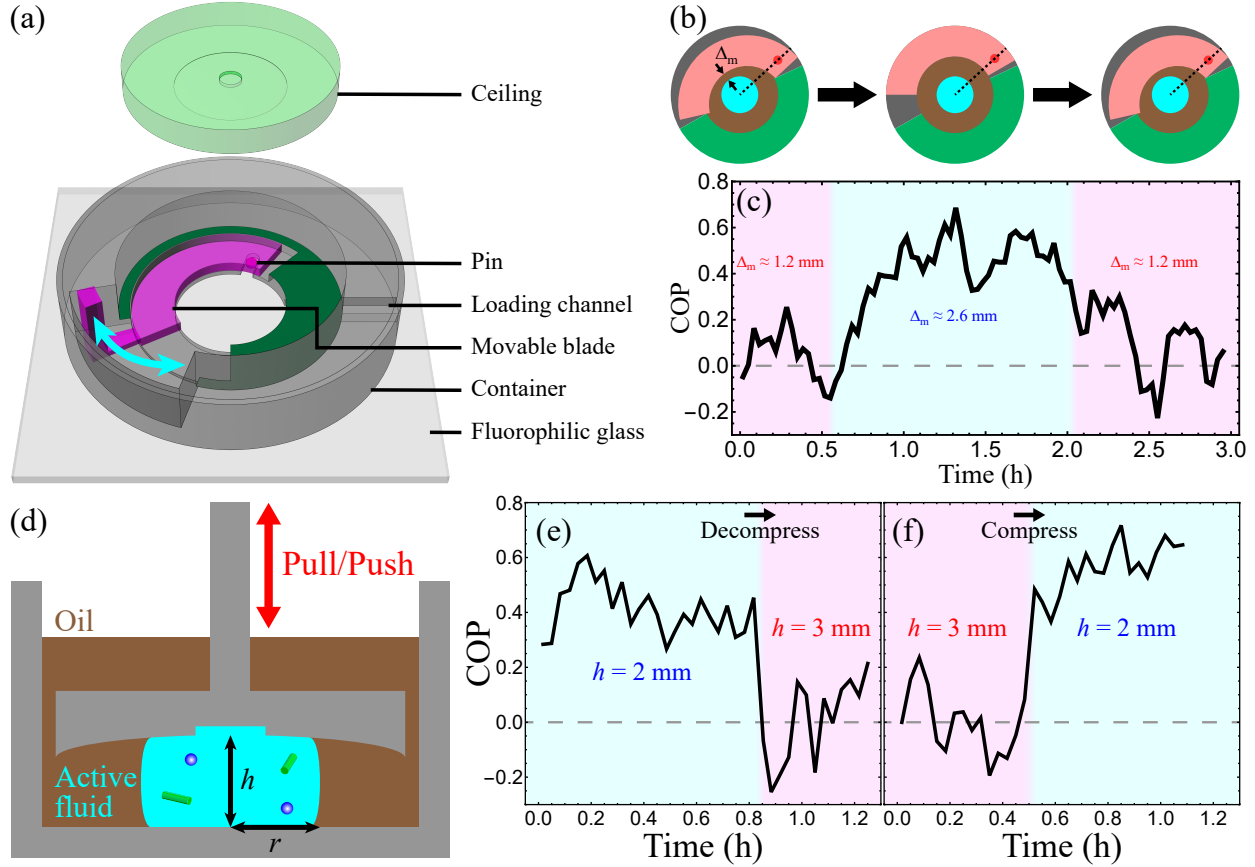

Fig. 8. We created millifluidic devices that trigger and suppress intradroplet circulatory flows in real time by manipulating the oil layer thickness (a–c) or deforming the droplet (d–f). (a) Novel millifluidic device with a movable blade (pink) that allows the amount of oil surrounding the droplet to be tuned. (b) Schematics of manipulating the oil layer thickness with a movable wall (pink). The wall was anchored to a cylindrical container (gray) by a pin (red dot) that could move radially (dashed black lines). Moving the pin toward the container center rotated the wall counterclockwise and reduced the minimum oil layer thickness ($\Delta_m$), whereas pushing the pin outward rotated the wall clockwise and increased the oil layer thickness. The corresponding 3D schematics are available in Supplementary Fig. S8. (c) The evolution of circulation order parameter (COP) of flows in a droplet ($r \approx 2.4$ mm, $h = 2$ mm) revealed that increasing the minimum oil layer thickness ($\Delta_m$) regulated the chaotic flows into circulatory flows (left pink to middle cyan areas). Conversely, decreasing the minimum oil layer thickness suppressed the circulatory flows (middle cyan to right pink areas). A video of manipulating the formation and deformation of intradroplet circulatory flows using the wall-movable device is available in Supplementary Video S3. (d) Schematic of compressing a droplet with a movable ceiling to manually tune the droplet height. Because the droplet volume was conserved, decreasing the droplet height $h$ enlarged the droplet radius $r$ and vice versa. The schematic of realizing such ceiling manipulation is shown in Supplementary Fig. S9. (e) Circulatory flows within the droplet were suppressed by decompressing the droplet (cyan to pink areas). The decompression reduced the droplet radius from $r \approx 2.0$ to 1.7 mm. (f) Conversely, compressing the droplet regulated the chaotic flows into circulatory flows (pink to cyan areas). The compression expanded the droplet radius from $r \approx$ 1.7 to 2.0 mm. A video of manipulating the formation and deformation of the circulatory flows with the ceiling-tunable device is available in Supplementary Video S4.

# Supplementary Information:
# Flow coupling between active and passive fluids across water–oil interfaces

Yen-Chen Chen[1], Brock Jolicoeur[2], Chih-Che Chueh[3], and Kun-Ta Wu[1,2,4,*]

[1]Department of Mechanical Engineering, Worcester Polytechnic Institute, Worcester, Massachusetts 01609, USA
[2]Department of Physics, Worcester Polytechnic Institute, Worcester, Massachusetts 01609, USA
[3]Department of Aeronautics and Astronautics, National Cheng Kung University, Tainan 701, Taiwan
[4]The Martin Fisher School of Physics, Brandeis University, Waltham, Massachusetts 02454, USA
[*]Corresponding: kwu@wpi.edu

## Table of Contents

**Supplementary Discussion S1: Influence of the ceiling geometry on intradroplet active fluid behavior**

Compressed active droplets self-propel,[1-4] which could prevent us from observing intradroplet flows for long durations with a fixed observation window. Therefore, we fixed the droplet by curving the ceiling of the fluidic device into a dome-like shape and then drilled a thin well at the dome center to reinforce droplet immobilization (Supplementary Fig. S1a). However, these shape modifications change the droplet shape, which might influence the self-organization of active fluids in droplets. Here, we examined how the intradroplet active fluid flows were influenced by the geometric parameters of the ceiling (i.e., the well dimensions and dome shape).

**Influence of well dimensions.** First, we investigated the role of the well geometry on intradroplet flows. The well was cylindrical; its geometry depended on its height (or depth) and radius. To examine how these parameters influenced intradroplet flows, we measured the circulation order parameter (COP) within a droplet while varying the well radius (Supplementary Fig. S2a) and depth (Supplementary Fig. S2b) separately. Across our explored parameters, our measured COPs remained steady (fluctuating between 0.4 and 0.6), which suggests that the well geometry did not play a significant role in the formation of intradroplet circulatory flows. In this study, we chose a well geometry (depth 0.2 mm, radius 1 mm) whose

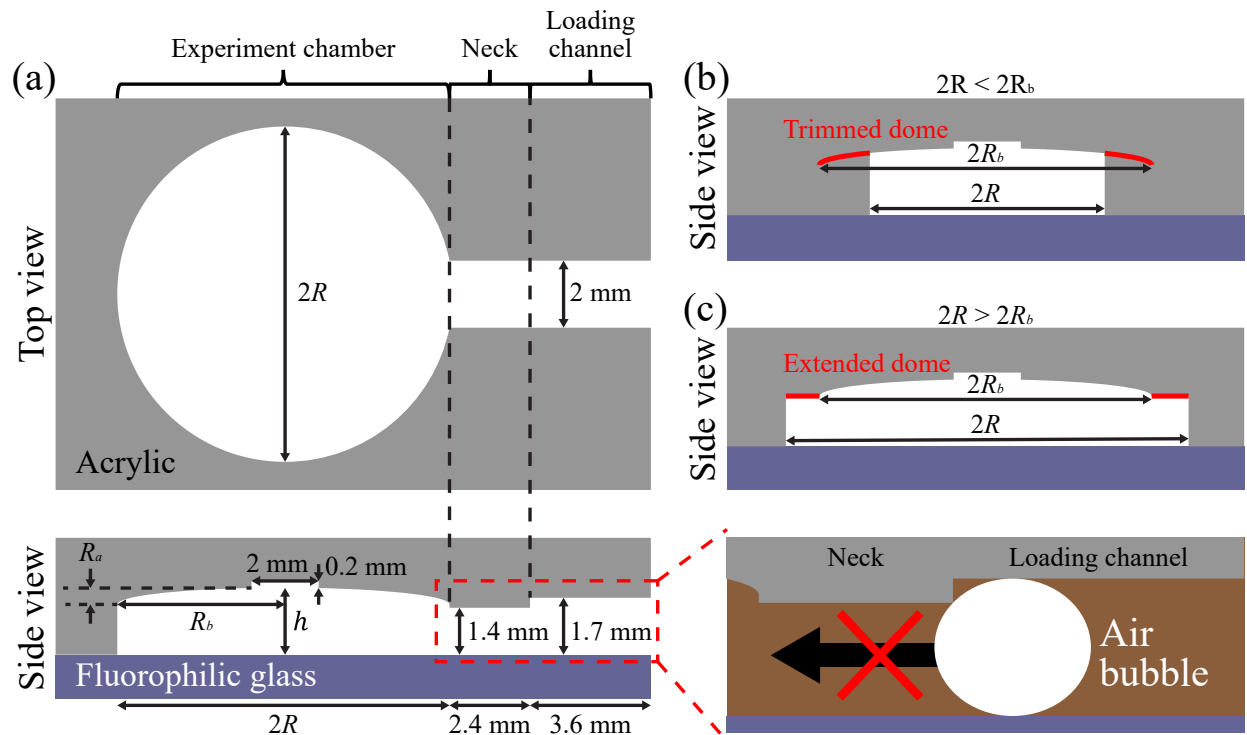

Supplementary Fig. S1: Millifluidic device for compressing an active fluid droplet. (a) The device contained a cylinder-like chamber to contain the oil and droplet, which were loaded via the loading channel. The channel was connected to the chamber via a neck to keep air bubbles out of the chamber (close-up). To fix the droplet in the chamber center, the chamber ceiling was curved into a half-oblate spheroidal shape whose semi-axes were $R_a$ and $R_b$, and a shallow well was drilled at the ceiling center. Close-up: An air-in-oil bubble in the loading channel could not spontaneously enter the chamber because of the smaller opening of the neck. (b) To fit a half-spheroidal dome into a smaller chamber ($2R < 2R_b$), the dome was trimmed (red curves). (c) Conversely, to match a dome to a larger chamber ($2R > 2R_b$), the dome was extended horizontally (red lines).

COP could not be distinguished from the case without a well (within error bars). We expected that using this well on the ceilings could reinforce droplet immobilization while minimizing the well's influence on intradroplet active fluid flows.

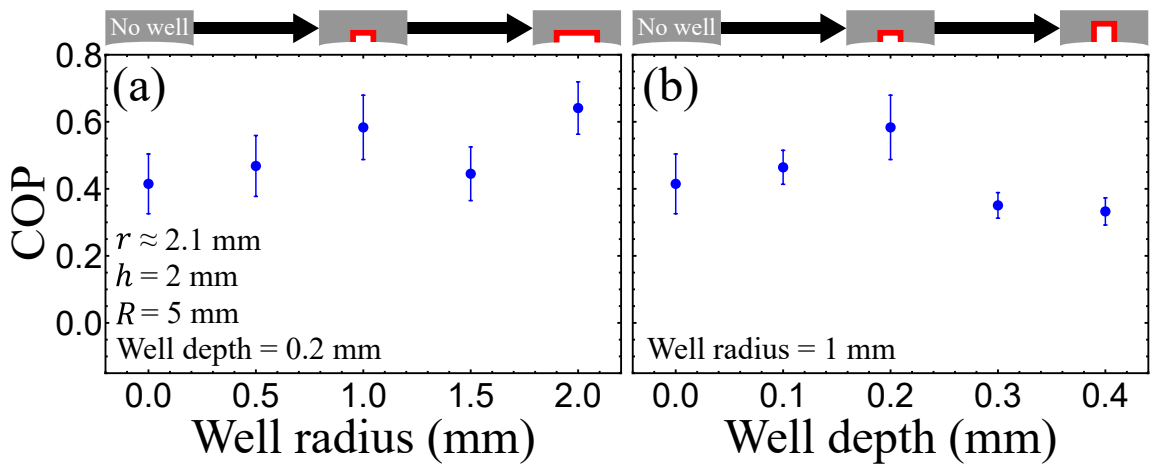

Supplementary Fig. S2: Intradroplet circulatory flows were not significantly influenced by the well dimensions. Geometries of the half-spheroidal domes were fixed ($R_a = 0.5$ mm, $R_b = 5$ mm). Varying either the well radius (panel a) or the well depth (panel b) did not significantly alter the circulation order parameter (COP), which suggests that the wells did not play a substantial role in intradroplet flows. Each error bar represents the standard deviation of the time-averaged COP.

**Influence of dome shape.** Next, we characterized how the intradroplet active fluid flows were influenced by the dome shape. The dome was shaped into a half-oblate spheroid whose geometry depended on its vertical and horizontal semi-axes (Supplementary Fig. S1). To characterize how the dome geometry influenced intradroplet circulatory flows, we systematically varied the vertical semi-axes $R_a = 0.25$–$1.5$ mm while maintaining the horizontal semi-axes $R_b = R = 5$ mm (Supplementary Fig. S1a) and then measured the corresponding COPs of intradroplet flows (Supplementary Fig. S3). Our measurements showed that circulatory flows persisted when the vertical semi-axes were shorter than ~1 mm (COP > 0.4); lengthening the vertical semi-axes longer than this limit weakened the formation of circulatory flows (COP ≤ 0.4). This result suggests that the dome shape influences the formation of intradroplet flows, but this influence was limited to largely curved dome ($R_a \gtrsim 1$ mm). To minimize the influence of the dome while immobilizing the droplets, we adopted half-spheroidal domes with a vertical semi-axis of $R_a = 0.25$–$0.5$ mm and a horizontal semi-axis of $R_b = 3.5$–$5$ mm.

In summary, observing the intradroplet active fluid flows over a long duration required compressing the droplets with a curved ceiling. However, the curved ceiling influenced the intradroplet fluid flows. To minimize this influence, we chose the ceiling shape that would not suppress development of circulatory flows but was sufficiently curved to fix the droplets. This arrangement provided a stationary active droplet that allowed us to investigate how the formation of intradroplet circulatory flows was controlled by other geometric parameters such as the droplet radius and oil layer thickness (Figs. 1–3).

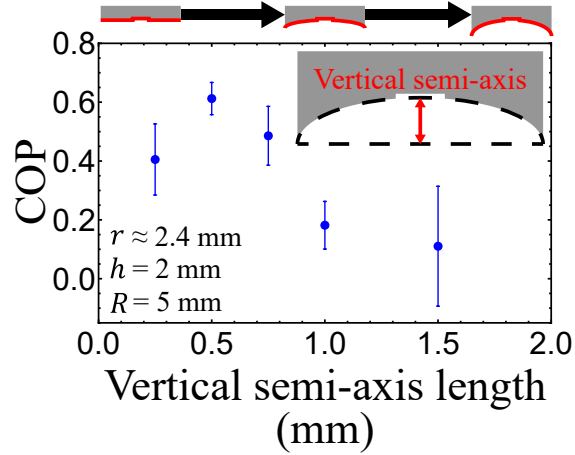

Supplementary Fig. S3: Intradroplet circulatory flows were influenced by the dome shape. The well geometry was fixed (radius 1 mm, depth 0.5 mm), and the domes were half-oblate spheroids with a horizontal semi-axis of 5 mm and vertical semi-axes ranging from 0.25 to 1.5 mm. Lengthening the vertical semi-axes increased the dome curvature. Such increases in curvature sharpened the droplet shape and suppressed development of circulatory flows when vertical semi-axes were longer than ~1 mm (circulatory order parameter [COP] ≤ 0.4); for the semi-axes shorter than this limit, circulatory flows were stable, which suggests that domes with a larger curvature had a greater influence on intradroplet flows. Thus, we selected a slightly curved dome ($R_a = 0.5$ mm, $R_b = 5$ mm) to compress the droplets. Each error bar represents the standard deviation of the time-averaged COP.

**Supplementary Discussion S2: Modeling active droplet systems with a continuum simulation**

To test whether an existing active fluid model was capable of describing the influences of active fluid–oil coupling on intradroplet active fluid flows,[3,5-25] we adopted the active droplet model established by Gao *et al.* because this model closely matched our experimental system which involved a microtubule-based active fluid in a water-in-oil droplet.[3] The model applied a cylindrical no-slip boundary (radius $R$, height $h$) with a curved ceiling, which was identical to the confinement geometry in the experiments (Fig. 1b). The boundary was filled with oil (density $\rho_o$, dynamic viscosity $\mu_o$) surrounding a concentric cylindrical active fluid (radius $r$, density $\rho_a$, dynamic viscosity $\mu_a$) modeled as a compressed water-in-oil droplet. The water–oil interface was modeled as variations of a phase function:

$$c = H\left(1 - \frac{\sqrt{x^2 + y^2}}{r}\right), \tag{Eq. S1}$$

where $H$ is the Heaviside step function with $c\left(\sqrt{x^2 + y^2} \leq r\right) = 1$ representing the active fluid and $c\left(\sqrt{x^2 + y^2} > r\right) = 0$ representing the oil. To speed up the simulations and to match our experimental arrangement for immobilizing the droplets, we fixed the water–oil interface so the phase function was kept constant, rather than evolving as in the Cahn–Hilliard model.[26] The phase function varied at the water–oil interface, which induced interfacial tension forces $\boldsymbol{T} \equiv (\gamma K/\epsilon)\boldsymbol{\nabla} c$, where $\gamma$ is the interfacial tension, $\epsilon$ is the interface thickness, and $K \equiv c(c-1)(c-1/2) - \epsilon^2 \nabla^2 c$ is the chemical potential that characterizes the phase variation within the interfacial region.[27] Within the interface, active fluid generated active stress ($\boldsymbol{\sigma}_a$) to induce self-driven flows ($\boldsymbol{u}$) in incompressible fluids ($\boldsymbol{\nabla} \cdot \boldsymbol{u} = 0$) that were governed by the Navier–Stokes equation:

$$\rho\left(\frac{\partial \boldsymbol{u}}{\partial t} + \boldsymbol{u} \cdot \boldsymbol{\nabla} \boldsymbol{u}\right) = \boldsymbol{\nabla} \cdot [-p\boldsymbol{I} + \mu(\boldsymbol{\nabla}\boldsymbol{u} + \boldsymbol{\nabla}\boldsymbol{u}^T)] + \boldsymbol{F}, \tag{Eq. S2}$$

where $\rho$ is the fluid density, $p$ is the fluid pressure, $\mu$ is the dynamic viscosity of the fluid, and $\boldsymbol{F} \equiv \boldsymbol{T} + \boldsymbol{\nabla} \cdot \boldsymbol{\sigma}_a$ is the net body force. Calculating the net body force required determining the active stress, which was exerted by the extensile microtubule bundles in the experiment. Here, we modeled each extensile bundle as a self-elongating rod whose center of mass is $\boldsymbol{x} = (x, y, z)$ orienting at $\boldsymbol{p} = (\sin\theta\cos\varphi, \sin\theta\sin\varphi, \cos\theta)$. The configurations of these rods were described with a mean-field probability distribution function, $\Psi(\boldsymbol{x}, \boldsymbol{p}, t)$, normalized as $\int_p \Psi d\boldsymbol{p} = 1$. To conserve probability, the distribution function satisfied the Smoluchowski equation as:

$$\frac{\partial \Psi}{\partial t} + \boldsymbol{\nabla} \cdot (\dot{\boldsymbol{x}}\Psi) + \boldsymbol{\nabla}_p \cdot (\dot{\boldsymbol{p}}\Psi) = 0, \tag{Eq. S3}$$

where $\boldsymbol{\nabla}_p \equiv \partial(\boldsymbol{I} - \boldsymbol{pp})/\partial \boldsymbol{p}$ is the surface derivative on the unit sphere.[10,12,28,29] Considering that these rods were only advected by fluid flows, the translational flux of the rods was determined as $\dot{\boldsymbol{x}} = \boldsymbol{u} - D_T \boldsymbol{\nabla} \ln \Psi$, where $D_T$ is the translational diffusion coefficient. The rotational flux induced by the fluid velocity gradient is $\dot{\boldsymbol{p}} = (\boldsymbol{I} - \boldsymbol{pp}) \cdot (\boldsymbol{\nabla}\boldsymbol{u} + 2\zeta \boldsymbol{D}) \cdot \boldsymbol{p} - D_R \boldsymbol{\nabla}_p \ln \Psi$, where $\zeta$ is the mean-field torque strength and $D_R$ is the rotational diffusion coefficient.[12,30-32] These equations describe the coupling between the fluid flows and the rods' translational and rotational distributions, but solving these equations was computationally expensive. To reduce the computation load, the Smoluchowski equation was coarse-grained as

$$\frac{\partial \boldsymbol{D}}{\partial t} - D_T \Delta \boldsymbol{D} + \boldsymbol{u} \cdot \boldsymbol{\nabla}\boldsymbol{D} = 4\zeta(\boldsymbol{D} \cdot \boldsymbol{D} - \boldsymbol{D}:\boldsymbol{S}) + \boldsymbol{\nabla}\boldsymbol{u} \cdot \boldsymbol{D} + \boldsymbol{D} \cdot \boldsymbol{\nabla}\boldsymbol{u}^T - 2\boldsymbol{E}:\boldsymbol{S} - 6D_R\left(\boldsymbol{D} - \frac{\boldsymbol{I}}{3}\right), \tag{Eq. S4}$$

where $E \equiv \frac{\nabla u + \nabla u^T}{2}$ is the strain rate tensor, $D \equiv \int_p pp\Psi dp$ is the orientational average of the second moment tensor of the rod orientation $pp$, and $S \equiv DD$.[29,33] The coarse-grained Smoluchowski equation allowed for determining the configuration of the self-elongating rods, which generated an active stress that was modeled to be proportional to the local rod orientational order:[21] $\sigma_a = \alpha cD$, where $\alpha$ is the activity coefficient ($\alpha > 0$ represents contracting rods and $\alpha < 0$ represents elongating rods) and $D \equiv \int_p pp\Psi dp$ represents the orientational order of the rods.[34] The active stress drove the fluids whose flows were determined via the Navier–Stokes equation (Eq. S2), which in return rearranged the rod configurations and reformulated the active stress via the coarse-grained Smoluchowski equation (Eq. S4). Both equations formed a feedback loop to simultaneously evolve the fluid flows, rod configurations, and active stresses.

To solve these equations and to develop a simulation platform for modeling our active droplet systems, we expressed both equations in explicit forms that were fed into COMSOL Multiphysics™, which solved the equations by the finite element method.[35-39] To feed in the Navier–Stokes equation (Eq. S2), we utilized the template of the 3D laminar flow model in the software. This template considers incompressible fluids ($\nabla \cdot u = 0$) governed by the Navier–Stokes equation with a net body force that can be expressed as

$$F = \begin{bmatrix} F_x \\ F_y \\ F_z \end{bmatrix} = \alpha \begin{bmatrix} c\left(\frac{\partial D_{xx}}{\partial x} + \frac{\partial D_{yx}}{\partial y} + \frac{\partial D_{zx}}{\partial z}\right) + D_{xx}\frac{\partial c}{\partial x} + D_{yx}\frac{\partial c}{\partial y} + D_{zx}\frac{\partial c}{\partial z} \\ c\left(\frac{\partial D_{xy}}{\partial x} + \frac{\partial D_{yy}}{\partial y} + \frac{\partial D_{zy}}{\partial z}\right) + D_{xy}\frac{\partial c}{\partial x} + D_{yy}\frac{\partial c}{\partial y} + D_{zy}\frac{\partial c}{\partial z} \\ c\left(\frac{\partial D_{xz}}{\partial x} + \frac{\partial D_{yz}}{\partial y} + \frac{\partial D_{zz}}{\partial z}\right) + D_{xz}\frac{\partial c}{\partial x} + D_{yz}\frac{\partial c}{\partial y} + D_{zz}\frac{\partial c}{\partial z} \end{bmatrix} + \frac{\gamma K}{\epsilon} \begin{bmatrix} \frac{\partial c}{\partial x} \\ \frac{\partial c}{\partial y} \\ \frac{\partial c}{\partial z} \end{bmatrix}, \quad \text{(Eq. S5)}$$

where $\frac{\partial c}{\partial x_i} = \frac{-x_i \delta\left(r - \sqrt{x^2 + y^2}\right)}{\sqrt{x^2 + y^2}}$, $K = c(c-1)\left(c - \frac{1}{2}\right) - \epsilon^2\left[\delta'\left(r - \sqrt{x^2+y^2}\right) - \frac{\delta\left(r - \sqrt{x^2+y^2}\right)}{\sqrt{x^2+y^2}}\right]$, and $\delta$ is the Dirac delta function. To include the coarse-grained Smoluchowski equation (Eq. S4), we rearranged the equation as:

$$\frac{\partial D}{\partial t} - D_T \Delta D + \begin{bmatrix} u_x \\ u_y \\ u_z \end{bmatrix} \cdot \nabla D = -\begin{bmatrix} a_{xx} & a_{xy} & a_{xz} \\ a_{yx} & a_{yy} & a_{yz} \\ a_{zx} & a_{zy} & a_{zz} \end{bmatrix} \circ D + \begin{bmatrix} f_{xx} & f_{xy} & f_{xz} \\ f_{yx} & f_{yy} & f_{yz} \\ f_{zx} & f_{zy} & f_{zz} \end{bmatrix}, \quad \text{(Eq. S6)}$$

where

$$a_{xx} = 4\zeta\left(D_{xx}^2 + D_{yy}^2 + D_{zz}^2 + 2D_{xy}D_{yx} + 2D_{xz}D_{zx} + 2D_{yz}D_{zy} - D_{xx}\right)$$
$$+ 2\left[(D_{xx} - 1)\frac{\partial u_x}{\partial x} + D_{yy}\frac{\partial u_y}{\partial y} + D_{zz}\frac{\partial u_z}{\partial z}\right] + (D_{xy} + D_{yx})\left(\frac{\partial u_x}{\partial y} + \frac{\partial u_y}{\partial x}\right)$$
$$+ (D_{xz} + D_{zx})\left(\frac{\partial u_x}{\partial z} + \frac{\partial u_z}{\partial x}\right) + (D_{yz} + D_{zy})\left(\frac{\partial u_y}{\partial z} + \frac{\partial u_z}{\partial y}\right) + 6D_R,$$

$$a_{yy} = 4\zeta\left(D_{xx}^2 + D_{yy}^2 + D_{zz}^2 + 2D_{xy}D_{yx} + 2D_{xz}D_{zx} + 2D_{yz}D_{zy} - D_{yy}\right)$$
$$+ 2\left[D_{xx}\frac{\partial u_x}{\partial x} + (D_{yy} - 1)\frac{\partial u_y}{\partial y} + D_{zz}\frac{\partial u_z}{\partial z}\right] + (D_{xy} + D_{yx})\left(\frac{\partial u_x}{\partial y} + \frac{\partial u_y}{\partial x}\right)$$
$$+ (D_{xz} + D_{zx})\left(\frac{\partial u_x}{\partial z} + \frac{\partial u_z}{\partial x}\right) + (D_{yz} + D_{zy})\left(\frac{\partial u_y}{\partial z} + \frac{\partial u_z}{\partial y}\right) + 6D_R,$$

$$a_{zz} = 4\zeta\left(D_{xx}^2 + D_{yy}^2 + D_{zz}^2 + 2D_{xy}D_{yx} + 2D_{xz}D_{zx} + 2D_{yz}D_{zy} - D_{zz}\right)$$
$$+ 2\left[D_{xx}\frac{\partial u_x}{\partial x} + D_{yy}\frac{\partial u_y}{\partial y} + (D_{zz}-1)\frac{\partial u_z}{\partial z}\right] + (D_{xy}+D_{yx})\left(\frac{\partial u_x}{\partial y} + \frac{\partial u_y}{\partial x}\right)$$
$$+ (D_{xz}+D_{zx})\left(\frac{\partial u_x}{\partial z} + \frac{\partial u_z}{\partial x}\right) + (D_{yz}+D_{zy})\left(\frac{\partial u_y}{\partial z} + \frac{\partial u_z}{\partial y}\right) + 6D_R,$$

$$a_{xy} = a_{yx} = 4\zeta\left(D_{xx}^2 + D_{yy}^2 + D_{zz}^2 + 2D_{xy}D_{yx} + 2D_{xz}D_{zx} + 2D_{yz}D_{zy} - D_{xx} - D_{yy}\right)$$
$$+ 2\left[\left(D_{xx}-\frac{1}{2}\right)\frac{\partial u_x}{\partial x} + \left(D_{yy}-\frac{1}{2}\right)\frac{\partial u_y}{\partial y} + D_{zz}\frac{\partial u_z}{\partial z}\right] + (D_{xy}+D_{yx})\left(\frac{\partial u_x}{\partial y} + \frac{\partial u_y}{\partial x}\right)$$
$$+ (D_{xz}+D_{zx})\left(\frac{\partial u_x}{\partial z} + \frac{\partial u_z}{\partial x}\right) + (D_{yz}+D_{zy})\left(\frac{\partial u_y}{\partial z} + \frac{\partial u_z}{\partial y}\right) + 6D_R,$$

$$a_{xz} = a_{zx} = 4\zeta\left(D_{xx}^2 + D_{yy}^2 + D_{zz}^2 + 2D_{xy}D_{yx} + 2D_{xz}D_{zx} + 2D_{yz}D_{zy} - D_{xx} - D_{zz}\right)$$
$$+ 2\left[\left(D_{xx}-\frac{1}{2}\right)\frac{\partial u_x}{\partial x} + D_{yy}\frac{\partial u_y}{\partial y} + \left(D_{zz}-\frac{1}{2}\right)\frac{\partial u_z}{\partial z}\right] + (D_{xy}+D_{yx})\left(\frac{\partial u_x}{\partial y} + \frac{\partial u_y}{\partial x}\right)$$
$$+ (D_{xz}+D_{zx})\left(\frac{\partial u_x}{\partial z} + \frac{\partial u_z}{\partial x}\right) + (D_{yz}+D_{zy})\left(\frac{\partial u_y}{\partial z} + \frac{\partial u_z}{\partial y}\right) + 6D_R,$$

$$a_{yz} = a_{zy} = 4\zeta\left(D_{xx}^2 + D_{yy}^2 + D_{zz}^2 + 2D_{xy}D_{yx} + 2D_{xz}D_{zx} + 2D_{yz}D_{zy} - D_{yy} - D_{zz}\right)$$
$$+ 2\left[D_{xx}\frac{\partial u_x}{\partial x} + \left(D_{yy}-\frac{1}{2}\right)\frac{\partial u_y}{\partial y} + \left(D_{zz}-\frac{1}{2}\right)\frac{\partial u_z}{\partial z}\right] + (D_{xy}+D_{yx})\left(\frac{\partial u_x}{\partial y} + \frac{\partial u_y}{\partial x}\right)$$
$$+ (D_{xz}+D_{zx})\left(\frac{\partial u_x}{\partial z} + \frac{\partial u_z}{\partial x}\right) + (D_{yz}+D_{zy})\left(\frac{\partial u_y}{\partial z} + \frac{\partial u_z}{\partial y}\right) + 6D_R,$$

$$f_{xx} = 4\zeta\left(D_{xy}D_{yx} + D_{xz}D_{zx}\right) + (D_{xy}+D_{yx})\frac{\partial u_y}{\partial x} + (D_{xz}+D_{zx})\frac{\partial u_z}{\partial x} + 2D_R,$$

$$f_{yy} = 4\zeta\left(D_{xy}D_{yx} + D_{yz}D_{zy}\right) + (D_{xy}+D_{yx})\frac{\partial u_x}{\partial y} + (D_{yz}+D_{zy})\frac{\partial u_z}{\partial y} + 2D_R,$$

$$f_{zz} = 4\zeta\left(D_{xz}D_{zx} + D_{yz}D_{zy}\right) + (D_{xz}+D_{zx})\frac{\partial u_x}{\partial z} + (D_{yz}+D_{zy})\frac{\partial u_y}{\partial z} + 2D_R,$$

$$f_{xy} = 4\zeta D_{xz}D_{zy} + D_{xx}\frac{\partial u_x}{\partial y} + D_{xz}\frac{\partial u_z}{\partial y} + D_{yy}\frac{\partial u_y}{\partial x} + D_{zy}\frac{\partial u_z}{\partial x},$$

$$f_{yx} = 4\zeta D_{yz}D_{zx} + D_{xx}\frac{\partial u_x}{\partial y} + D_{zx}\frac{\partial u_z}{\partial y} + D_{yy}\frac{\partial u_y}{\partial x} + D_{yz}\frac{\partial u_z}{\partial x},$$

$$f_{xz} = 4\zeta D_{xy}D_{yz} + D_{xx}\frac{\partial u_x}{\partial z} + D_{xy}\frac{\partial u_y}{\partial z} + D_{yz}\frac{\partial u_y}{\partial x} + D_{zz}\frac{\partial u_z}{\partial x},$$

$$f_{zx} = 4\zeta D_{yx}D_{zy} + D_{xx}\frac{\partial u_x}{\partial z} + D_{yx}\frac{\partial u_y}{\partial z} + D_{zy}\frac{\partial u_y}{\partial x} + D_{zz}\frac{\partial u_z}{\partial x},$$

$$f_{yz} = 4\zeta D_{xz}D_{yx} + D_{yx}\frac{\partial u_x}{\partial z} + D_{yy}\frac{\partial u_y}{\partial z} + D_{xz}\frac{\partial u_x}{\partial y} + D_{zz}\frac{\partial u_z}{\partial y}, \text{ and}$$

$$f_{zy} = 4\zeta D_{xy}D_{zx} + D_{xy}\frac{\partial u_x}{\partial z} + D_{yy}\frac{\partial u_y}{\partial z} + D_{zx}\frac{\partial u_x}{\partial y} + D_{zz}\frac{\partial u_z}{\partial y}.$$

The rearranged equation was fed into the software by means of the built-in stabilized convective diffusion equation. Solving these equations numerically required defining the system domains and the associated boundary conditions. As such, we imported 3D computer-aided designs (SOLIDWORKS) identical to the geometries of our experimental containers (Fig. 1b) as system domains and then imposed to the domain surfaces a no-slip boundary condition: $\boldsymbol{u} = \boldsymbol{0}$. Because the rods were limited within the domain, we also imposed a no-flux boundary condition: $\boldsymbol{n} \cdot \nabla \Psi = 0$, or equivalently $\boldsymbol{n} \cdot \nabla \boldsymbol{D} = \boldsymbol{0}$, where $\boldsymbol{n}$ represents the unit vectors normal to domain surfaces, without enforcing rod orientations at boundaries.[29,33] To evolve the fluid flows and rod configurations, we initialized the fluids as quiescent fluids ($\boldsymbol{u} = \boldsymbol{0}$) under uniform pressure ($p = 0$) with uniformly suspended isotropic rods whose translational and orientational distributions were perturbed with 15 random modes:[10,11]

$$\Psi(\boldsymbol{x}, \boldsymbol{p}, 0) = \frac{1}{4\pi}\left[1 + \sum_{i=1}^{15} \epsilon_i P_i(\boldsymbol{p}) \cos(\boldsymbol{k}_i \cdot \boldsymbol{x} + \varsigma_i)\right], \quad \text{(Eq. S7)}$$

which determined the initial second moment tensor as

$$\boldsymbol{D}(\boldsymbol{x}, 0) = \int_{\boldsymbol{p}} \boldsymbol{pp}\Psi(\boldsymbol{x}, \boldsymbol{p}, 0)d\boldsymbol{p} = \frac{\boldsymbol{I}}{3} + \frac{1}{4\pi}\sum_{i=1}^{15} \epsilon_i \cos(\boldsymbol{k}_i \cdot \boldsymbol{x} + \varsigma_i) \int_{\boldsymbol{p}} \boldsymbol{pp}P_i(\boldsymbol{p})d\boldsymbol{p}, \quad \text{(Eq. S8)}$$

where $\epsilon_i$ is the random numbers in $[-0.01, 0.01]$, $\boldsymbol{k}_i$ is the random wave numbers whose components are random numbers in $[\pi, 15\pi]$ mm$^{-1}$, $\varsigma_i$ is the random phases in $[0, 2\pi]$, and $P_i(\boldsymbol{p}) = \sum_{j=1}^{3}\sum_{k=1}^{4} \xi_{ijk} g_k^j$ is the random polynomials of sine and cosine with $\xi_{ijk}$ as the random numbers in $[0, 1]$ and $g_k$ defined as: $g_1 \equiv \sin(\theta)$, $g_2 \equiv \cos(\theta)$, $g_3 \equiv \sin(\varphi)$, and $g_4 \equiv \cos(\varphi)$. The initialized second moment tensor and fluid flow were evolved for 3 hours with the selected model parameters (Supplementary Table S1). The resulting flow field ($\boldsymbol{u}$) was analyzed to determine the circulation order parameters and flow profiles (Fig. 6), which were compared with experimental outcomes (Figs. 3&5).

| Symbol | Description | Value |
|---|---|---|
| $\alpha$ | Activity coefficient ($\alpha < 0$: elogating rods; $\alpha > 0$: contracting rods) | –50 Pa |
| $\gamma$ | Water–oil interfacial tension[40] at 25 °C | 0.072 N/m |
| $\epsilon$ | Thickness of the water–oil interface | 50 μm |
| $D_T$ | Translational diffusion coefficient of rods | 0.011 m²/s |
| $D_R$ | Rotational diffusion coefficient of rods | 10 s$^{-1}$ |
| $\zeta$ | Mean-field torque strength on rods induced by flow velocity gradient | 100 s$^{-1}$ |
| $\rho_a$ | Density of active fluid (96% water) at 25 °C | 997 kg/m³ |
| $\rho_o$ | Density[41] of oil (hydrofluoroether, 3M Novec 7500) at 25 °C | 1614 kg/m³ |
| $\mu_a$ | Dynamic viscosity of active fluid (96% water) at 25 °C | 0.00089 Pa·s |
| $\mu_o$ | Dynamic viscosity[41] of oil (hydrofluoroether) at 25 °C | 0.00124 Pa·s |

Supplementary Table S1: Parameters used in the simulations. The activity coefficient, translational diffusion coefficient, and rotational diffusion coefficient were selected to match the flow speeds in the simulation with those in the experiments (~10 μm/s). The mean-field torque strength was selected to be four times larger than the rotational diffusion coefficient ($\zeta > 4D_R$) to enforce strong flow alignment of the rods. The water–oil interfacial tension was approximated as the surface tension of water[40] at 25 °C.

**Supplementary Discussion S3: Shear stress coupling across water–oil interface**

Flow coupling between active fluid and oil is the consequence of dynamic interactions between the fluids across their interfaces. To advance our understanding of these interactions, we used the simulation to investigate shear stress coupling across interfaces between water (inactive fluid) and oil. We used our established active fluid simulation (Fig. 6), turned off activity of active fluid ($\alpha = 0$), imposed shear stress within the droplet ($r = 2.4$ mm, $h = 2$ mm) by arranging a concentric cylinder (radius 1.2 mm) that rotated at a constant angular velocity ($\omega = 0.015$ s$^{-1}$) with a surface azimuthal velocity of 18 μm/s, and then evolved the fluid flows until they reached the steady state (Supplementary Fig. S4a). To reveal the role of oil layer thickness in shear stress coupling, we repeated the simulation for various oil layer thickness ($\Delta = 0.23$–9.6 mm) and analyzed the profiles of azimuthal velocity (Supplementary Fig. S4b) and corresponding magnitude of shear stress $\tau \equiv \left|\langle \mu\rho \frac{d}{d\rho}\left(\frac{v_\theta}{\rho}\right)\rangle\right|$, where $\mu$ is viscosity of either water or oil depending on the radial coordinate, $\rho$, and $\langle \ \rangle$ indicates averaging over $\theta$ (Supplementary Fig. S4c). Our analysis showed that the profiles of azimuthal flow depended on oil layer thickness because the imposed no-slip boundary condition at the outer boundary enforced the flow to decay to quiescence at the outer boundary; as such, a thinner oil layer drove the flow to decay more quickly (Supplementary Fig. S4b). Conversely, the profiles of shear stress were nearly independent of oil layer thickness because the stress is permitted to be nonzero at the outer boundary (Supplementary Fig. S4c). The thickness-independent profiles showed that the shear stresses induced within the droplets penetrated the water–oil interface and decayed in oil with a universal millimeter–scale decay length ($L_s \approx 1$ mm). This universal decay length scale of shear stress is consistent with flow coupling length scale observed in our experiments and model (Figs. 3–7), which implies that the flow coupling between active fluid and oil is related to the stress coupling between these two fluids. In active fluid droplets, the shear stress was induced by extensile microtubule bundles; our simulation suggests that this active shear stress could penetrate water–oil interface into the oil to a millimeter–scale depth. When the oil was deeper than this depth ($\Delta > L_s$), the dynamics of extensile bundles were only coupled to a portion of oil near interface; conversely, when the oil was thinner than this depth ($\Delta \lesssim L_s$), the bundle dynamics were coupled to the entire oil layer. Thus, in our experiments and model, we observed that the intradroplet circulatory flows could develop when the thickness of the oil layer was sufficiently large ($\Delta > L_s$) but could be suppressed by the active fluid–oil coupling when the thickness of the oil layer became smaller ($\Delta \lesssim L_s$) (Figs. 3&6).

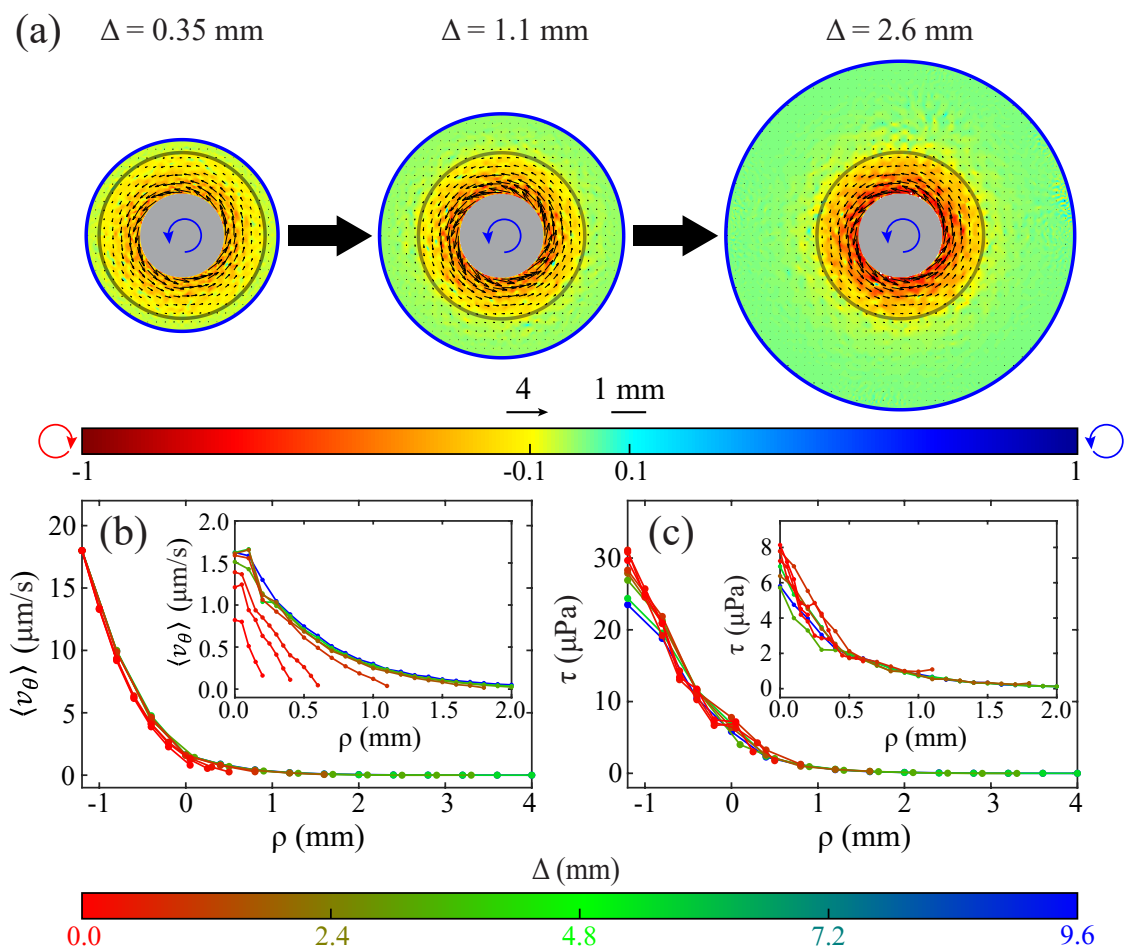

Supplementary Fig. S4: Simulation results show that shear stresses on both sides of water–oil interfaces were coupled with a millimeter–scale coupling length. (a) Velocity fields and vorticity maps (plotted as in Fig. 1c) of three identical water-in-oil droplets ($r = 2.4$ mm, $h = 2$ mm) surrounded by oil layers of different thicknesses. Within each droplet was a concentric rotating cylinder (angular velocity 0.015 s$^{-1}$) that generated intradroplet shear stress. (b) Profile of azimuthal velocities for droplets immersed in oil layer thicknesses from 0.23 to 9.6 mm. The horizontal axis represents the radial axis in cylindrical coordinate with the origin shifted to the droplet interface (Fig. 5c inset). Inset: Close-up near the water–oil interface ($\rho = 0$). (c) Profile of corresponding shear stress magnitude $\tau \equiv \left| \langle \mu \rho \frac{d}{d\rho} \left( \frac{v_\theta}{\rho} \right) \rangle \right|$. Inset: Close-up near the water–oil interface.

**Supplementary Discussion S4: The role of interfacial properties on flow coupling**

We have investigated how the flow coupling between active fluid and oil across the water–oil interfaces influences the self-organization of intradroplet active fluid. As this coupling is across interfaces, it is expected to be influenced by interfacial properties. We used our established simulation platform to investigate how the coupling and intradroplet active fluid flows were affected by two interfacial properties: viscosity contrast and interfacial tension.

**Viscosity contrast.** To reveal the role of viscosity contrast in the fluid dynamics of the active droplet system (Fig. 6a), we varied the oil viscosity from 0.00089 to 0.124 Pa·s and the viscosity constrast, $\bar{\mu} \equiv \eta_o/\eta_w$, from 1 to 140 and analyzed the time-averaged circulation order parameter (COP; Supplementary Fig. S5a). Our analysis showed that in the simulation of the droplet ($r = 2.4$ mm, $h = 2$ mm) with the thicker oil layer ($\Delta = 2.6$ mm) that supported circulatory flows, viscosity contrast had no significant effect on COP except for the lowest viscosity contrast ($\bar{\mu} = 1$) where active fluid and oil had an identical viscosity (blue dots), whereas the noncirculating droplet immersed in the thinner oil layer ($\Delta = 1.1$ mm) developed circulation as the oil became more viscous (red dots). This result showed that the interfacial viscosity contrast promoted the formation of intradroplet circulatory flows.

To gain deeper insight into this viscosity contrast-aided circulation, we analyzed the cross-correlation length between the active fluid and oil (Supplementary Fig. S5c) and found that the correlation lengths in both droplets decayed with viscosity contrast. This decay indicates that increasing the viscosity in oil enhanced energy dissipation, reduced the range of oil that active fluid could drive, and thus suppressed the flow coupling between active fluid and oil. With the suppressed flow coupling, the formation of intradroplet circulation would be dominated by droplet geometry; our previous studies showed that the development of intradroplet circulation requires droplet geometry with an aspect ratio lower than 3, i.e. $r/h \lesssim 3$.[42] The droplets in our simulation had an aspect ratio of $r/h = 1.2$, which met the criteria, so intradroplet circulatory flows were better supported in a system with higher viscosity contrast where flow coupling was suppressed and droplet geometry became the prime factor for circulation development (red dots in Supplementary Fig. S5a).

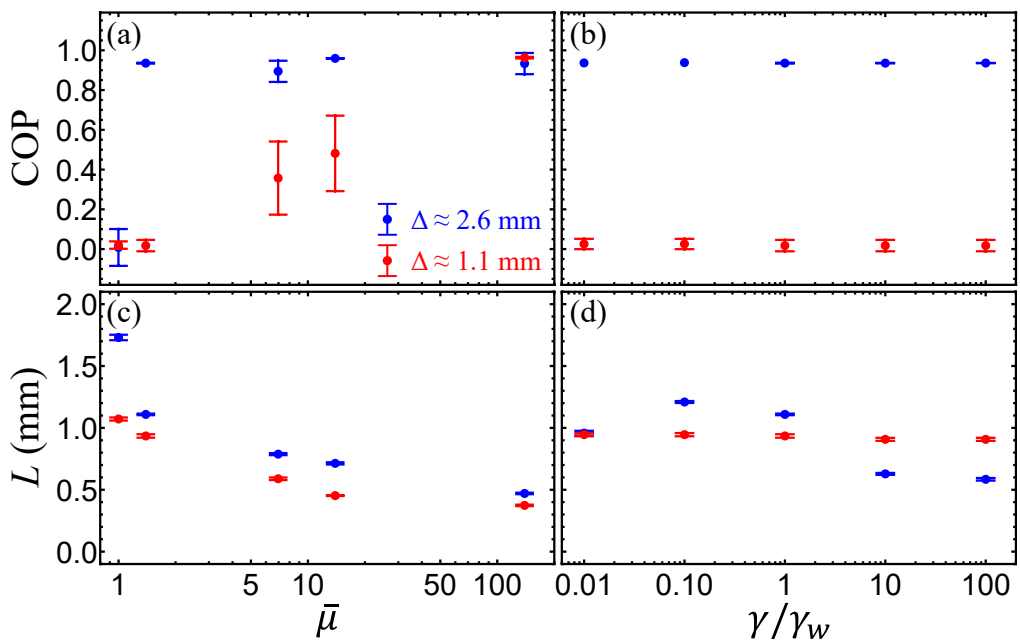

Supplementary Fig. S5: A simulation comparing the role of viscosity contrast and interfacial tension in intradroplet flows and active fluid–oil flow coupling for droplets ($r = 2.4$ mm, $h = 2$ mm) immersed in a thick oil layer (blue dots; $\Delta = 2.6$ mm) or a thin oil layer (red dots; $\Delta = 1.1$ mm). The *y*-axes are time-averaged circulation order parameter (COP) and cross-correlation length ($L$) and the *x*-axes are viscosity contrast, $\bar{\mu} \equiv \mu_o/\mu_w$, and relative interfacial tension, $\gamma/\gamma_w$, where $\gamma_w = 0.072$ N/m is the surface tension of water at 25 °C.[40] Error bars in panels a and b represent standard deviation and error bars in panels c and d represent uncertainty in fitting the correlation function to an exponential function (Fig. 7a inset). The simulation showed that increasing viscosity contrast promoted the formation of intradroplet circulatory flows (panel a) because it suppressed active fluid–oil flow coupling (panel c), whereas interfacial tension played nearly no role in intradroplet flows and flow coupling (panels b&d).

**Interfacial tension.** Interfacial tension controls how a droplet will deform in response to external force so as to influence the active fluid–oil coupling and associated intradroplet flows. In our simulation, we neglected this deformation by imposing onto droplets a condition that fixed the droplet geometry, so we hypothesize that in our simulation, the flow coupling and intradroplet active fluid flows will not be affected by a change in interfacial tension. To test this hypothesis within the simulation, we adopted the same pair of droplet systems (Fig. 6a), varied the interface tension $\gamma$ from 0.00072 to 7.2 N/m or relative interfacial tension $\gamma/\gamma_w$ from 0.01 to 100 where $\gamma_w = 0.072$ N/m is surface tension of water at 25 °C, and analyzed the time-averaged COP (Supplementary Fig. S5b). Our analyses showed that the COP did not vary with interfacial tension, which indicates that the intradroplet active fluid flows were not influenced by interfacial tension. To test if the interfacial tension affected the flow coupling between active fluid and oil, we analyzed the cross-correlation length between the active fluid and oil (Supplementary Fig. S5d), revealing that in both droplet systems, the cross-correlation length remained nearly unchanged (within ~20% variation) as we varied the interfacial tension. These analyses demonstrated that the interfacial tension played no role in flow coupling and intradroplet active fluid flows. However, this result was the consequence of assuming a shape-fixed droplet. Experimental results have shown that the droplet deforms more easily as the interfacial tension is lowered,[43] so varying interfacial tension is expected to influence active fluid–oil flow coupling and intradroplet flows. Investigating the role of interfacial tension in active droplet system requires experiments that reduce interfacial tension so that the interfacial fluctuation can be observed, and simulations where the droplet deformation is permitted.[3,44]

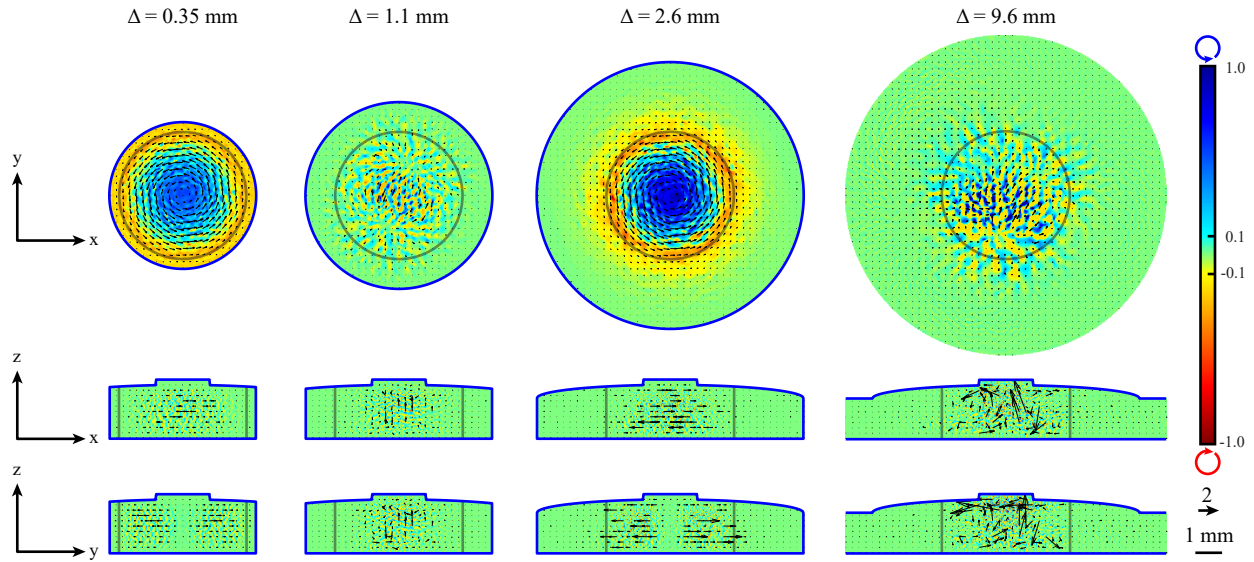

Supplementary Fig. S6: Simulated instantaneous cross-sectional velocity fields and vorticity maps in active droplets that had the same geometry ($r = 2.4$ mm, $h = 2$ mm) but were immersed in oil layers of different thicknesses ($\Delta =0.35$–$9.6$ mm). The velocity fields and vorticity maps were plotted as in Fig. 1c. The columns from left to right represent various oil layer thicknesses. (The fourth column plots show only the central portion [12 mm wide] of the system.) The rows from top to bottom represent cross-sections at the $xy$ midplane ($z = h/2$), $xz$ midplane ($y = 0$), and $yz$ midplane ($x = 0$), with the axis origin at the bottom center of each droplet. These plots show not only that oil layer thickness influenced intradroplet active fluid flow, but also that when the droplet is in the circulation state, the fluid mainly flows horizontally, whereas when the droplet is in the chaotic state, the fluid flows both horizontally and vertically.

**Supplementary Discussion S5: Formation of two-dimensional nematic layers at water–oil interfaces**

Microtubule-based active fluid systems are known to deposit microtubules onto the water–oil interface and form 2D active nematics.[45-53] Therefore, we expected microtubules to gather at the water–oil interface and develop 2D active nematics. To verify this expectation, we prepared a compressed water-in-oil active droplet ($r \approx 2.4$ mm, $h = 1$ mm, $\Delta \approx 2.4$ mm) and imaged the microtubules at the water–oil interface of the droplet bottom with fluorescence microscopy (there is a ~1-μm layer of oil between the droplet and the glass plate[48]; Supplementary Fig. S7a). The microtubules were labeled with Alexa 647, which can be imaged with a Cy5 filter cube (excitation: 618–650 nm, emission: 670–698 nm, Semrock, 96376) (Supplementary Fig. S7b&c). The images showed that the microtubules formed a layer of nematics at the water–oil interface with multiple motile plus- and minus-half defects.[16,23,24,48-50,52-55] These motile microtubule-based defects served as a dynamic boundary that confined the active fluid. To examine whether such a dynamic boundary was coupled to the self-organization of the confined active fluid, we simultaneously imaged microtubules at the bottom interface and at the droplet midplane for 15 minutes and then analyzed microtubule motion with the particle image velocimetry algorithm to extract the velocity fields of the microtubule motions (arrows in Supplementary Fig. S7b).[56] The velocity fields enabled us to analyze the circulation order parameter (COP) as a function of time (Supplementary Fig. S7d). Our analyses showed that microtubules at the midplane developed circulatory flows (COP ≈ 0.5, solid blue curve in Supplementary Fig. S7d), whereas at the bottom interface the microtubule flows were chaotic (|COP| ≲ 0.2, dashed blue curve in Supplementary Fig. S7d). This comparison indicated that the microtubule flows in bulk and at the interface were not coupled. To confirm such bulk–interface decoupling, we repeated the experiments but immersed the droplet in a thinner oil layer (Δ ≈ 1.1 mm, Supplementary Fig. S7c) that did not support circulatory flows in the droplet bulk (|COP| ≲ 0.2, solid red curve in Supplementary Fig. S7d), and we found that the microtubule motion at the interface remained chaotic (|COP| ≲ 0.2, dashed red curve in Supplementary Fig. S7d). Our analyses showed that microtubule motion at the interface of the droplet bottom were not related to microtubule motions in bulk. This work demonstrates that microtubules developed 2D active nematics at the water–oil interfaces of droplets and that nematic motions at the droplet bottom interfaces were decoupled from flows in the droplet bulk.

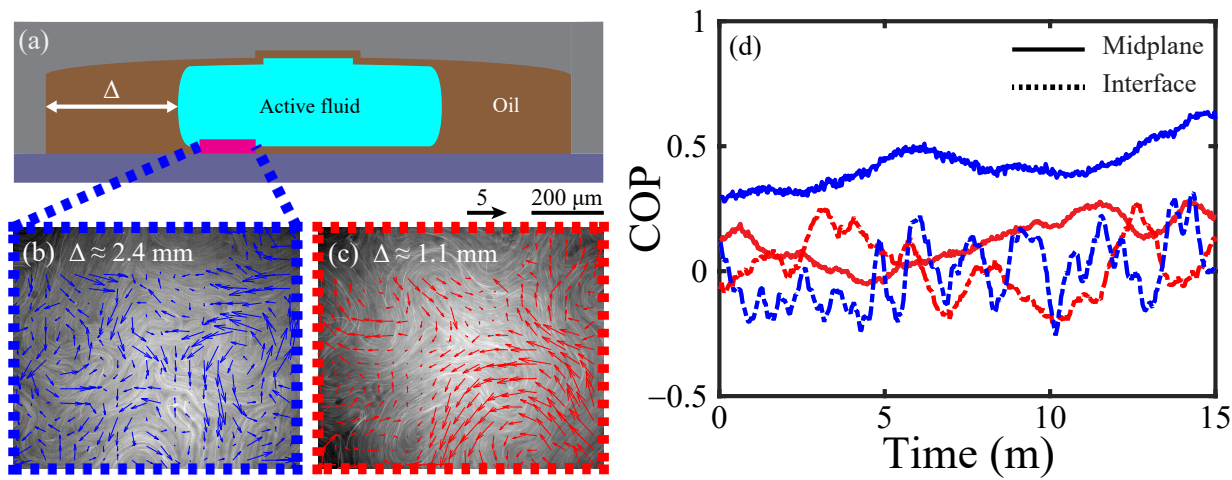

Supplementary Fig. S7: At water–oil interfaces at the bottom of the droplet, microtubules self-organized into a layer of motile nematics whose flows were not influenced by microtubule flows in the droplet bulk. (a) Schematic of imaging the nematic layer at the water–oil interface at the droplet bottom (magenta line). (b&c) Fluorescent micrographs of the microtubule-based nematic layer at the water–oil interfaces of droplets that had the same shape ($r = 2.4$ mm, $h = 1$ mm) but were immersed in oil layers of different thicknesses ($\Delta$). The arrows are normalized velocity fields of corresponding instant nematic flows. (d) Evolution of circulation order parameter (COP) of microtubule flows at the midplane (solid curves) and at water–oil interface (dashed curves). Blue curves represent the droplet with a thick oil layer ($\Delta \approx 2.4$ mm) that developed circulatory flows in bulk (COP ≈ 0.5, solid blue curve) and red curves represent the droplet with a thinner oil layer ($\Delta \approx 1.1$ mm) that developed chaotic flows in bulk (COP≈ 0, solid red curve). However, regardless of how the microtubules flowed in bulk, the microtubule-based nematics flowed chaotically at the interface at the bottom of the droplet (dashed curves).

However, our analyses did not suggest that the dynamic boundary was decoupled from the self-organization of intradroplet flows. We have shown that oil near the lateral interface flowed faster when the active fluid flowed chaotically than when the active fluid developed circulatory flows (Fig. 5d), which implies that the microtubule motion at the droplet lateral interface was coupled to the intradroplet fluid flows. Unveiling such a coupling would require further studies monitoring the microtubules at the lateral droplet interfaces.

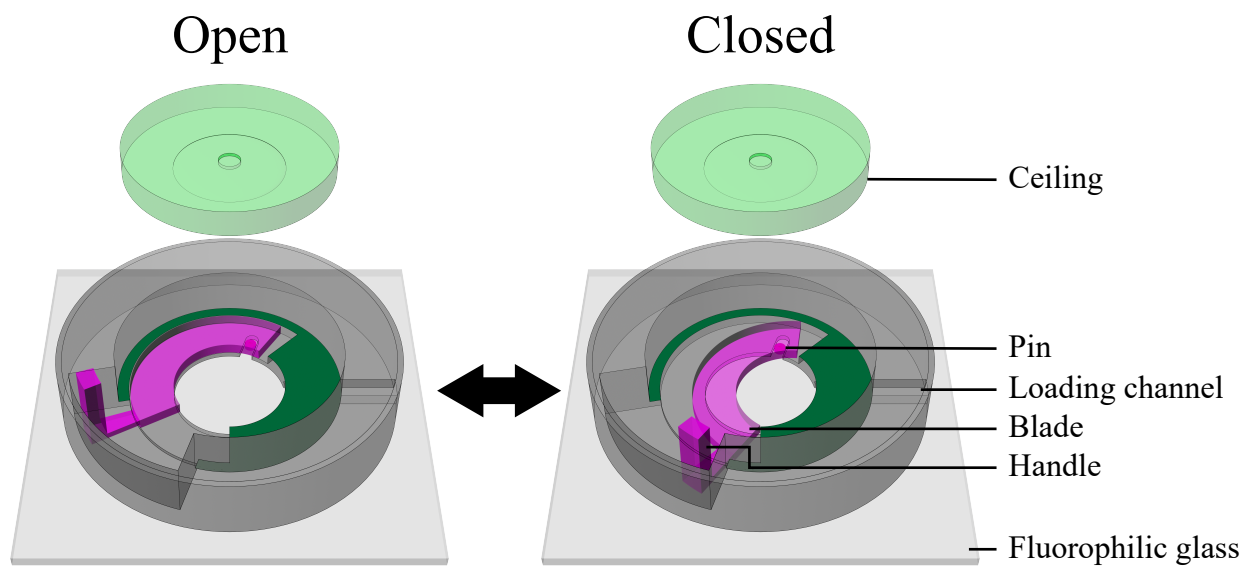

Supplementary Fig. S8: Schematics of the wall-movable milli-fluidic device used to manipulate the layer thicknesses of the oil that immersed an active droplet. The device consisted of a container (gray) that had a cylindrical chamber (radius $R$ = 5 mm, height 1.5 mm), a curved ceiling ($R_a = 0.5$ mm, $R_b = R = 5$ mm, light green) that sat on the dark-green platform to enclose the chamber, and a blade (pink) that altered the oil layer thickness. The blade comprised a blade body whose inner side wall served as a movable boundary of the chamber, a pin that was constrained in a groove of the container, and a handle used to manually rotate the blade body around the pin during experiments. Rotating the blade counterclockwise slid the pin inward, shifted the blade midpoint toward chamber center by 1.4 mm, and thus shrank the chamber (left to right). Conversely, rotating the blade clockwise slid the pin outward, shifted the blade midpoint away from the chamber center, and expanded the chamber (right to left). This device enabled real-time tuning of the layer thickness of the oil that immersed the droplets.

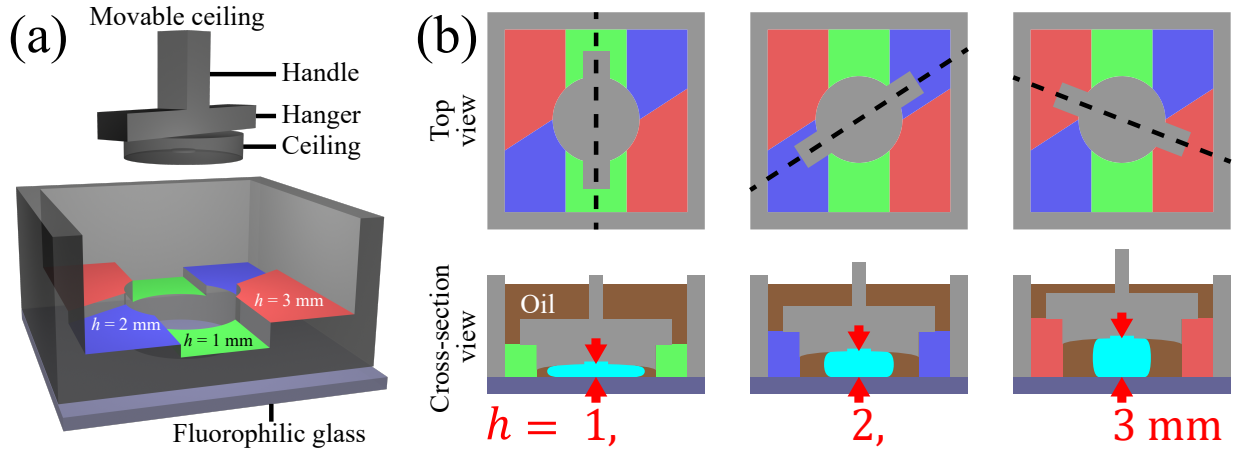

Supplementary Fig. S9: Millifluidic device with movable ceiling for manually compressing a droplet. (a) The device contains a cylindrical well (radius $R = 5$ mm) with a movable ceiling. The ceiling consists of a handle used to manually adjust the ceiling height, a curved ceiling ($R_a = 0.5$ mm, $R_b = R = 5$ mm) used to fix the droplet at the ceiling center, and a hanger that was designed to sit on the colored platforms (green, blue, and red) to place the ceiling at various heights ($h$). To ensure that the droplet remained in oil while moving the ceiling, we submerged the central well, ceiling, and platforms in oil that was held in a cubical container (gray walls). The front side of the container was transparentized to visualize the platform design. (b) The ceiling hanger was placed on green, blue, and red platforms, causing the ceiling to compress the water-in-oil droplet (cyan) with the heights of $h$ = 1, 2, and 3 mm, respectively. The dashed lines indicate cross-section planes.

Supplementary Video S1: Circulation of microtubule-based active fluids confined in a water-in-oil droplet. The droplet was compressed into a cylinder-like shape with a height of 2 mm and a radius of 1 mm. The time stamp is hour:minute:second.

Supplementary Video S2: The oil layer thickness controlled the formation of circulatory flows in an active fluid droplet, shown by tracer movements. The droplet was compressed into a cylinder-like shape with a height of 2 mm and a radius of 2.4 mm and immersed in an oil bath. When the oil had a layer thickness of 2.6 mm, circulatory flows developed. Decreasing the oil layer thickness to 1.1 mm suppressed the circulatory flows. The time stamp is hour:minute:second.

Supplementary Video S3: A novel millifluidic device manipulated circulatory flows within a water-in-oil active fluid droplet without directly contacting the droplet. The device has one movable wall. Circulatory flows developed after the blade was moved away from the droplet, which increased the oil layer thickness from 1.2 to 2.6 mm (00:33:54–00:34:01). The circulatory flows faded away after the blade approached the droplet, which decreased the oil layer thickness from 2.6 to 1.2 mm (02:00:51–02:00:54). The time stamp is hour:minute:second.

Supplementary Video S4: A novel millifluidic device with a movable ceiling controlled formation of intradroplet circulatory flows. The circulatory flows were suppressed by lifting the ceiling from 2 to 3 mm (00:50:18–00:50:38), whereas the circulatory flows were triggered by lowering the ceiling from 3 to 2 mm (00:30:20–00:30:52). During the times when the ceiling was being moved, the image brightness was oversaturated by the room light that was needed to manually lower or lift the ceiling. The time stamp is hour:minute:second.